\documentclass[aps,prd,reprint,nofootinbib,superscriptaddress,showpacs,showkeys]{revtex4-1}
\usepackage[usenames,dvipsnames]{xcolor}
\usepackage{amsmath,amssymb,amsfonts}
\usepackage{hyperref}
\usepackage{graphicx}
\usepackage[utf8]{inputenc}
\usepackage{graphicx}
\usepackage{slashed}
\usepackage{color}
\usepackage{bm,bbm}
\usepackage{makecell}
\usepackage[normalem]{ulem}

\usepackage{tikz}
\usetikzlibrary{shapes.geometric}

\definecolor{darkgreen}{rgb}{0,0.35,0}

\newcommand{\nwse}[3]{\ensuremath{#1^{#2}_{\phantom{#2} #3}}}

\newcommand{\um}{\mathbbm{1}}

\definecolor{yellowish}{RGB}{253,250,203}
\definecolor{blueish}{RGB}{203,203,253}

\newcommand{\cecs}{Centro de Estudios Científicos (CECs) Av.~Arturo Prat~514, Valdivia, Chile}
\newcommand{\uab}{Universidad Andrés Bello, Av.~República~440, Santiago, Chile}
\newcommand{\udec}{Departamento de Física, Universidad de Concepción, Casilla 160-C, Concepción, Chile}
\newcommand{\ulbisi}{Physique Théorique et Mathématique, Université Libre de Bruxelles and International Solvay Institutes, Campus Plaine C.P.~231, B-1050 Bruxelles, Belgium}
\newcommand{\unc}{Departamento de Física, Universidad Nacional de Colombia, Bogotá, Colombia}

\numberwithin{equation}{section}
\begin{document}

\title{Supersymmetric 3D model for gravity with $SU(2)$ gauge symmetry, mass generation and effective cosmological constant}

\author{Pedro D. Alvarez}
\email{alvarez@cecs.cl}
\affiliation{\cecs}

\author{Pablo Pais}
\email{pais@cecs.cl}
\affiliation{\cecs}
\affiliation{\uab}
\affiliation{\ulbisi}

\author{Eduardo Rodríguez}
\email{eduarodriguezsal@unal.edu.co}
\affiliation{\unc} 

\author{Patricio Salgado-Rebolledo}
\email{pasalgado@udec.cl}
\affiliation{\cecs}
\affiliation{\ulbisi}
\affiliation{\udec}

\author{Jorge Zanelli}
\email{z@cecs.cl}
\affiliation{\cecs}


\begin{abstract}
A Chern--Simons system in $2+1$ dimensions invariant under local Lorentz rotations, $SU(2)$ gauge transformations, and local $\mathcal{N}=2$ supersymmetry transformations is proposed. The field content is that of $(2+1)$-gravity plus an $SU(2)$ gauge field, a spin-1/2 fermion charged with respect to $SU(2)$ and a trivial free abelian gauge field. A peculiarity of the model is the absence of gravitini, although it includes gravity and supersymmetry. Likewise, no gauginos are present. All the parameters involved in the system are either protected by gauge invariance or emerge as integration constants. An effective mass and effective cosmological constant emerge by spontaneus breaking of local scaling invariance. The vacuum sector is defined by configurations with locally flat Lorentz and $SU(2)$ connections sporting nontrivial global charges. Three-dimensional Lorentz-flat geometries are spacetimes of locally constant negative --or zero--, Riemann curvature, which include Minkowski space, AdS$_3$, BTZ black holes, and point particles. These solutions admit different numbers of globally defined, covariantly constant spinors and are therefore good candidates for stable ground states. The fermionic sector in this system could describe the dynamics of electrons in graphene in the long wavelength limit near the Dirac points, with the spin degree of freedom of the electrons represented by the $SU(2)$ label. If this is the case, the $SU(2)$ gauge field would produce a spin-spin interaction giving rise to strong correlation of electron pairs.
\end{abstract}


\maketitle

\tableofcontents

\section{Introduction}            
Supersymmetry (SUSY) is the only nontrivial symmetry that mixes internal and spacetime transformations. SUSY is a compelling idea could solve many puzzles in particle physics and cosmology. It imposes severe restrictions on particle multiplets and makes it possible to establish positive energy and stability theorems. Additionally, it improves renormalizability of otherwise unrenormalizable theories, provides a protection mechanism for the hierarchy of the observed energy scales in unified theories, and even suggests candidates for dark matter (for a modern introduction to SUSY, see, e.g., ~\cite{Mar11}).

Standard SUSY models describe particle multiplets in which fermions and bosons come in equal numbers and have the same mass. Since this pairings are experimentally observed, it means that SUSY must be severely broken at currently accessible energy scales. Whether SUSY is restored at a higher energy scale is yet to be seen, but decades of experimental searches have tightened the constraints on the simplest plausible SUSY models, if not discarded them completely \cite{CMS13a,CMS13b,ATL14a,ATL14b}.

As far as it can be experimentally assessed, fermions and bosons play very different roles in nature: bosons are carriers of interactions whereas fermions constitute matter currents that interact via the exchange of gauge bosons. All bosons, with the sole exception of the Higgs, are spin-one fields in the adjoint representation of a gauge group, mathematically described by one-form connections in a fiber bundle. Fermions, on the other hand, are spin-1/2 fields in a vector representation of the same gauge group, mathematically described as zero-form sections in the bundle. These distinct features seem hard to reconcile with SUSY, which transforms fermions into bosons and vice~versa. But that is not necessarily so. There is a class of SUSY models where fermions and bosons are combined in a gauge connection, in which fermions are sections and bosons are connections~\cite{Alv11,Alv13}. Models of this type have a long tradition of precursors that dates back to the pioneering work of Deser, Jackiw, and Templeton, Ach\'{u}carro and Townsend \cite{Des82a,Des82b,Ach86,Ach89,Cha90, Ba96,Tro97,Tro98}.

When supersymmetry is locally realized, gravity is naturally included \cite{Van81} and brings in the gravitino, the spin-3/2 superpartner of the metric~\cite{Cre78}. The novelty of the system we discuss here is that the local realization of SUSY combines fermions and bosons in a connection for a superalgebra that does not require the presence of gravitini. One of the simplest three-dimensional version of this model contains an abelian gauge field, a spin-1/2 Dirac fermion, a spin connection and the vielbein combined into a connection for $osp(2|2)$. The action is taken as the Chern--Simons (CS) form for this connection and all parameters in the action (up to an overall constant) are fixed by gauge invariance \cite{Alv11,Alv13}. The mass of the fermion emerges as an integration constant --not necessarily zero--, while the bosons remain massless. It is the breaking of a built-in Weyl symmetry what provides a mass gap, rather than the breaking of SUSY. This action can describe (massive) electrons moving in a curved graphene background including topological defects and external electromagnetic fields \cite{Iorio11}.

A similar construction can be attempted in $3+1$ dimensions by extending the gauge algebra $SO(3,1) \times U(1)$ to $usp(2,2|1)$, which includes supersymmetry. Although the connection is uniquely defined along the lines of the three-dimensional case instead of a CS form, which do not exist in even dimensions, the natural Lagrangian is the Yang--Mills form that gives ries to an action of the MacDowell--Mansouri type \cite{MacDowell:1977jt}. This breaks the $USP(2,2|1)$-invariance and the system is invariant under local $SO(3,1) \times U(1) \subset USP(2,2|1)$ transformations, which is expected in any low energy description of gravity \cite{Stelle:1979aj,McCarthy:1985nt,Wise:2006sm}. The resulting theory includes gravity with a positive cosmological constant, Maxwell electrodynamics and a charged spin-1/2 field minimally coupled to both gravity and electromagnetism, plus a Nambu--Jona-Lasinio self-coupling term \cite{Alv13}. In this case, although supersymmetry is broken, its underlying presence uniquely determines both the field content and the interactions in the action. The exercise can be repeated in higher dimensions and for different gauge groups. The general situation is this: in odd $D$ the CS form allows to build supersymmetric models, while for even $D$ the Yang--Mills construction leads necessarily to scenarios with broken SUSY.

Graphene near the Dirac points is characterized by electrons described by a two-component massless Dirac field, where each component corresponds to the states occupying the two triangular sublattices of the hexagonal graphene lattice. Therefore, the spinor index of the electron field is not related to the orientation of the spin. The spin degree of freedom of the electron must be represented by an additional $SU(2)$ index. In this way, the system acquires a local symmetry, the freedom to choose the quantization of axis independently at each point in the manifold --or at each lattice site. This $SU(2)$ symmetry had already been used in the context of the Jordan--Wigner transformation that maps localized fermions into collective bosonic states and vice~versa in the Hubbard model \cite{HZ93}.

In this work we consider a $2+1$ model which includes gravity, an $su(2)$ gauge field $A^I$, and two spin-1/2 fermion fields $\psi_i$. Gravitation is represented by the dreibein $e^a_\mu$ ---the soldering form between the tangent space and the spacetime manifold--- and the Lorentz connection $\omega^a{}_{b\mu}$.  All fields combine into a one-form gauge connection $\mathbb{A}$ transforming in the adjoint representation of $su(2,1\vert 2)$. Supersymmetry is locally realized, yet there are no gravitini and therefore this is not standard supergravity. As explained in detail in section~\ref{action}, this feature comes about because the gravitino is replaced by the combination $\nwse{e}{a}{\mu} \gamma_a \psi_i$, where $\psi$ is a zero-form, spin-1/2 Dirac fermion and $\nwse{e}{a}{\mu}$.

Having established the action and the field equations for the model, in Section~\ref{sec:vacsol} we turn to the study of vacuum solutions, i.e., configurations with $\psi_i=0$, for which both the Lorentz and the $su(2)$ curvatures vanish. Interestingly enough, as shown in ~\cite{Alv14a}, AdS$_3$ satisfies these conditions. Here we show that the three-dimensional black hole, endowed with an $su(2)$-singlet charge $W$, is also a vacuum solution. The solution can also be extended to the rotating case. These 2+1 black holes, like the original one \cite{Ban92}, are locally ``pure gauge'' but globally nontrivial.  In Section~\ref{sect-charges}, direct application of Noether's theorem yields the conserved charges.

In Section~\ref{killingspinors}, we address the question of stability by establishing the existence of Killing spinors. By direct integration of the Killing spinor equation, we prove the existence of globally-defined Killing spinors for particular values of the mass $M$ and $W$. This means that these solutions are Bogomolny–Prasad–Sommerfield states (BPS, \cite{Bogomolny:1975de,Prasad:1975kr}) and therefore good ground state candidates. We close in Section~\ref{summary} with conclusions and an outlook for future work.

\section{The system} \label{action}  

Graphene, the two-dimensional hexagonal lattice of carbon atoms, was predicted by Wallace \cite{Wallace} and had been studied by various authors in the last sixty years \cite{Semenoff}.  This system attracted considerable attention in recent years, after the discovery  by  Geim and Novoselov of an ingenious method for its production in the laboratory, which has allowed the study of its exceptional physical properties \cite{Novoselov2004}.

Here we consider a system describing a two-dimensional spatial manifold where a spin-1/2 Dirac field propagates interacting with the geometry of the manifold in a manner that electrons in graphene would do, via the minimal coupling to the spin connection in a Weyl invariant manner. We adopt the point of view that although the kinematics of graphene takes place in a (2+1)-dimensional spacetime, the fermions describe electrons in 3+1 dimensions. This means, in particular, that the fermions in this system belong to a spin-1/2 irreducible representation of $SU(2)$. The invariance under the changes in the local definition of spin quantization axis introduces a corresponding $SU(2)$ gauge symmetry that brings about an interaction between the spin and the gauge field.

It turns out that this system consisting of the charged spin-1/2 field, the geometry of the 2+1 spacetime, the electromagnetic interaction and the $SU(2)$ gauge field, can be neatly packaged as a CS gauge theory for the $su(2,1| 2)$ algebra in three-dimensional spacetime. This is the smallest superalgebra containing $su(2)$ and the Lorentz algebra $so(1,2) \equiv sl(2;\mathbb{R}) \equiv sp(2;\mathbb{R})$ in three-dimensional spacetime.  Additionally, there are complex supersymmetry generators and a $u(1)$ generator. An explicit $4 \times 4$ supermatrix representation of this superalgebra is given in Appendix~\ref{App1}.

\subsection{Connection and curvature}    
We take a connection one-form in the algebra spanned by the Lorentz generators $\mathbb{J}_a$, the $su(2)$ generators $\mathbb{T}_I$, the supercharges $\mathbb{Q}_i$ and $\overline{\mathbb{Q}}^{i}$, and a central $u(1)$ extension $\mathbb{Z}$. As usual the coefficients are the dynamical fields,
\begin{equation}  \label{A}
\mathbb{A}=\omega^a \mathbb{J}_a + A^I \mathbb{T}_I + \overline{\mathbb{Q}}^{i} \gamma_a e^a\psi_i +\overline{\psi}^{i} e^a\gamma_a \mathbb{Q}_i+b\mathbb{Z},
\end{equation}
where $\omega^a=\frac{1}{2}\epsilon^a{}_{bc}\omega^{bc}$ is the Lorentz (spin) connection,\footnote{We use the convention $\epsilon_{012}=+1$.} $A^I$ is the $su(2)$ connection and $e^a$ is the dreibein one-form (local frame). Flat space indexes take values $a=0,1,2$ in $so(1,2)$ and we will mostly omit spinor indexes $\alpha=1,2$. The index $I=1,2,3$ is in the adjoint and $i=1,2$ in the fundamental representation of $SU(2)$. The two-component complex spinor $\psi_i^\alpha$ is a zero-form, and the field $b$ is an abelian one-form. The spinor need not be charged with respect to $b$ because $\mathbb{Z}$ is central.

In order to incorporate a fermion matter field in the connection, we use a soldering form so the two fields combine into a one-form. The local $SO(1,2)$ frame combines correctly with the spacetime index of the $\gamma$-matrices respecting Lorentz symmetry. This combination also gives correct volume forms to build Lagrangian densities in curved space. The \emph{composite} field $\xi_\mu^\alpha=\gamma_\mu\psi^\alpha$ has identically vanishing spin-3/2 component ($\mu$ are curved space indexes), so it is not the Rarita-Schwinger field. As we show below, neither gauginos nor gravitini are present. Thus, it is clear that the theory constructed with the connection $\mathbb{A}$ is fundamentally different from supergravity and from the usual supersymmetric formulations of CS theories in three dimensions \cite{Ach86,Ach89}. Similar constructions have also been considered in \cite{Krasnov:2011pp,TorresGomez:2012sr}.

Part of the geometric structure and the details of the representation can be readily seen from the curvature two-form, $\mathcal{F}\equiv d\mathbb{A}+\mathbb{A}^2$,\footnote{Exterior product of forms is understood and wedges are therefore omitted, except where ambiguities might arise. We also omit internal indexes whenever it does not lead to ambiguities, so that $\overline{\psi}\psi \equiv{\overline{\psi}}^i\psi_i$ and $\overline{\psi}\sigma^I\psi \equiv{\overline{\psi}}^i(\sigma^I)_i^{\ j}\psi_j$.}
\begin{equation}
\mathcal{F}=\mathcal{F}^a \mathbb{J}_a+\mathcal{F}^I \mathbb{T}_I + {\overline{\mathbb{Q}}}^i\mathcal{F}_i+{\overline{\mathcal{F}}}^i\mathbb{Q}_i+\mathcal{F}_{(b)} \mathbb{Z}\,,
\end{equation}
whose components are given by
\begin{align}
\mathcal{F}^a &=R^a-\epsilon^a_{\ bc}e^be^c\overline{\psi}\psi\,, \label{F^a}\\
\mathcal{F}^I &=F^I-i\epsilon_{abc}e^a e^b\overline{\psi}\gamma^c\sigma^I\psi\,, \\
\mathcal{F}_i &= D_i^{\ j}(\slashed{e}\psi_j)\,, \\
{\overline{\mathcal{F}}}^i &=-({\overline{\psi}}^j\slashed{e})\overleftarrow{D}_j^{\ i}\,, \\
\mathcal{F}_{(b)} &= db -i\epsilon_{abc}e^a e^b\overline{\psi}\gamma^c\psi\, .\label{Fb}
\end{align}
Here $R^a=d\omega^a+\frac{1}{2}\epsilon^a_{\ bc}\omega^b \omega^c$ is the Lorentz curvature 2-form, and $F^I = dA^I + \frac{1}{2} \epsilon^I{}_{JK}A^J A^K$ is the $su(2)$ curvature. We denote by $\slashed{e}$ the contraction $e^a\gamma_a =\gamma_\mu dx^\mu$, and use $D$ for the exterior covariant derivative for an $so(1,2)\times su(2)$ connection. In particular, for the spin-$1/2$ fundamental representation,
\begin{align}
D_i^{\ j}&=\delta^j_i d+\frac{1}{2}\delta^j_i\omega^a\gamma_a-\frac{1}{2} A^I(\sigma_I)_i^{\ j}\,, \label{D}\\
\overleftarrow{D}_i^{\ j}&=\delta^j_i \overleftarrow{d}-\frac{1}{2}\delta^j_i\omega^a\gamma_a+\frac{1}{2}A^I(\sigma_I)_i^{\ j}\,,\nonumber
\end{align}
and $\Omega^m\overleftarrow{d}=(-1)^m d\Omega^m$ for an $m$-form. In (\ref{D}) we have used the full Lorentz connection (metric compatible) $\omega^a$. The Lorentz connection can be split uniquely as $\omega^a=\mathring{\omega}^a+\kappa^a$, where $\mathring{\omega}^a$ is the torsion-free connection and $\kappa^a$ is the contorsion 1-form, so that $de^a + \epsilon^a_{\ bc}\mathring{\omega}^b e^c \equiv 0$ and $T^a = \epsilon^a_{\ bc}\kappa^b e^c$.

In three dimensions, spinors have mass-dimension one and the vielbein have dimensions of length, so that the combination $e^a \psi$ is dimensionless. This cancellation is a reflection of a scale invariance, consequence of the trick used to incorporate spin-1/2 matter firlds in the connection (\ref{A}). This means that the following Weyl transformations
\begin{equation}
e^a\rightarrow \lambda(x)e^a, \qquad \psi\rightarrow \lambda(x)^{-1}\psi \,,
\end{equation}
is a symmetry of the classical action. The components of the curvatures, (\ref{F^a})-(\ref{Fb}), are sourced by scale-invariant combinations of the local frames and matter. In Section \ref{sec:vacsol}, we will discuss classical solutions for the background geometry that provide a mechanism for spontaneous breaking of scale invariance.

\subsection{Action}     
In three dimensions, we choose as gauge (quasi-) invariant Lagrangian the CS form, $L_{\text{CS}}(\mathbb{A})\equiv \frac{\kappa}{2}\langle \mathbb{A}d \mathbb{A}+\frac{2}{3}\mathbb{A}^3 \rangle$, where the bracket $\left\langle \cdots \right\rangle$ stands for the supertrace in the corresponding representation. In this way the Lagrangian containing the dynamical fields is
\begin{equation}  \label{L0}
L_0(\omega^a,e^a,A^I,b,\psi)=L_{\text{CS}}(\mathbb{A})\,,
\end{equation}
where the relevant non-vanishing bilinear supertraces are
\begin{align}
\langle \mathbb{J}_a \; \mathbb{J}_b \rangle &=\frac{1}{2}\eta_{ab}\,, \\
\langle \mathbb{T}_I \; \mathbb{T}_J \rangle &= \frac{1}{2} \delta_{IJ}\,,\\
\langle \mathbb{Q}_i^{\alpha}\; {\overline{\mathbb{Q}}}_{\beta}^j \rangle &=-\delta^\alpha_\beta\delta_i^j.
\end{align}
The fact that all traces involving $\mathbb{Z}$ vanish means that $b$ is decoupled from the Lagrangian and no field equation involves it.
\footnote{Since $\mathbb{Z}$ is a central extension, the $U(1)$ field can be trivially included in the system by adding its CS form $bdb$, plus the corresponding minimal coupling to the fermion, $\overline{\psi}\slashed{b}\psi$. We ignore this possibility here in order to keep the discussion as simple as possible; the coupling to the $U(1)$ field was discussed in \cite{Alv11}.} The general properties of the CS form guarantee that the Lagrangian (\ref{L0}) changes at most by a closed form (boundary term) under local $su(2,1|2)$ transformations. The case of a manifold with boundary will be discussed in the  next section.

Expanding the Lagrangian (\ref{L0}) yields
\begin{equation}
L_0=L_{\text{CS}}(\omega^a)+L_{\text{CS}}(A^I)+ d^3x |e|\mathcal{L}_\psi\,,
\end{equation}
where $|e|=\det [e^a_{\ \mu}]$,  $L_\text{CS}(\omega^a)$ is the Lagrangian of CS gravity \cite{Witten:1988hc}, $L_\text{CS}(A^I)$ is the CS form for the $su(2)$ connection $A^I$ and $d^3x |e| \mathcal{L}_\psi =\frac{\kappa}{2}\overline{\psi}^i\slashed{e}(\overleftarrow{D}_i{}^j-D_i{}^j)\slashed{e}\psi_j$.
The fermionic Lagrangian can be written as the minimal coupling contribution plus an extra non-minimal coupling with the torsion
\begin{equation}\label{Lpsi}
\mathcal{L}_\psi=\kappa \ \overline{\psi}[\gamma^\mu D_\mu-\overleftarrow{D}_\mu \gamma^\mu+\frac{1}{2}\epsilon_a{}^{bc}T^a{}_{bc}]\psi\,,
\end{equation}
where $T^a{}_{bc}\equiv E_a{}^\nu E_b{}^\lambda T^a{}_{\nu\lambda}$ are the components of the torsion two-form, $T^a\equiv De^a=(1/2) T^a{}_{\mu \nu} dx^\mu \wedge dx^\nu$. In (\ref{Lpsi}) the derivatives act only on the spinors, while the torsion arises from the derivatives acting on the vielbein. In terms of the torsion-free connection the coefficient of the coupling with the torsion is changed,
\begin{equation}
\mathcal{L}_\psi= \kappa \ \overline{\psi}[\gamma^\mu \mathring{D}_\mu-\overleftarrow{\mathring{D}}_\mu \gamma^\mu+\frac{1}{4}\epsilon_a{}^{bc}T^a{}_{bc}]\psi\,.
\end{equation}
In either form, the fermion couples with completely antisymmetric part of the torsion contracted with the epsilon tensor.

We define the Dirac adjoint as
\begin{equation}
 \overline{\psi}^i_\alpha=i\psi^{\dagger\beta}_j C_{\beta\alpha}\delta^{ji}\,,
\end{equation}
or simply as $\overline{\psi}=i\psi^\dagger C$. Here
\begin{equation}
 (\gamma_a)^\dagger=C\gamma_a C\,, \quad C^\dagger=-C\,, \quad C^2=-1\,.
\end{equation}
so that $C^\dagger C=1$. We use gamma matrices $\{\gamma_a,\gamma_b \}=2\eta_{ab}$, with $[\eta_{ab}]=\text{diag}(-,+,+)$. The fermionic covariant bilinears $\overline{\psi}\psi$ and $\overline{\psi}\sigma_I\psi$ are real, and $\overline{\psi}\gamma_a\psi$ $\overline{\psi}\gamma_a \sigma_I\psi$ are imaginary. In three dimensions $\gamma_{ab}\sim\gamma_a$ so this list is complete. From this we see that all the gauge fields, the Dirac Lagrangian and the whole action are real.

\textbf{Equations of motion.} The field equations are given by the zero-curvature conditions,
\begin{equation}
\mathcal{F}^a =0\,, \quad \mathcal{F}^I =0\,,  \label{eom1}
\end{equation}
the Dirac equation and the resulting equation from the variation with respect to the local frame.

Varying with respect to $\overline{\psi}$ and dropping a boundary term $\partial_\mu(-\kappa \ |e|\delta\overline{\psi}\psi)$, we obtain the Dirac equation,
\begin{equation}\label{eompsi}
0=2[\gamma^\mu\mathring{D}_\mu+m]\psi+\epsilon_{abc}\mathring{\omega}^a_{\ \mu}E^{b\mu}\gamma^c\psi +|e|^{-1}\partial_\mu(|e| \gamma^\mu)\psi \,,
\end{equation}
where the last two terms are required by hermiticity. In (\ref{eompsi}) we have defined the quantity,
\begin{equation}\label{mass}
 m=\frac{1}{8}\epsilon^{abc}T_{abc}\,,
\end{equation}
associated to a nonzero torsion of the spacetime.

Varying with respect to the local frame gives $\delta(|e| \mathcal{L}_\psi)=|e|\delta e^a_{\ \mu}\tau_a^{\ \mu}+\kappa \ \partial_\mu[|e|\epsilon^{\mu\nu}_{\ \ a}\delta e^a_{\ \nu} \overline{\psi}\psi]$, where
\begin{align}
 \tau_a^{\ \mu}=&\kappa \left[ \overline{\psi}(E_a^{\ \mu}\gamma^b D_b-\overleftarrow{D}_bE_a^{\ \mu}\gamma^b)\psi \nonumber \right.\\
 &\left.- \overline{\psi}(\gamma^\mu D_a-\overleftarrow{D}_a \gamma^\mu)\psi+ \frac{1}{3}E_a^{\ \mu}\epsilon^{bcd}T_{bcd}\overline{\psi}\psi\right]\,.
\end{align}
Therefore, the field equation
\begin{equation}
 \tau_a^{\ \mu}=0\,,
\end{equation}
guarantees that the stress-energy tensor,
\begin{align}\label{tmunu}
 t^{\mu\nu}=&\kappa \ g^{\mu\nu}[\overline{\psi}(\gamma^\rho D_\rho-\overleftarrow{D}_\rho\gamma^\rho)\psi+\frac{1}{3}\epsilon^{abc}T_{abc}\overline{\psi}\psi] \nonumber\\
 &-\kappa \ \overline{\psi}[\gamma^{(\mu}D^{\nu)}-\overleftarrow{D}^{(\mu}\gamma^{\nu)}]\psi\,,
\end{align}
vanishes on-shell \footnote{Here $t^{\mu\nu}\equiv(2/\sqrt{-g})\delta(\sqrt{-g} \mathcal{L}_\psi)/\delta g_{\mu\nu}
= (1/2)\eta^{ab}(E_a^{\ \mu}\tau_b^{\nu}+E_a^{\ \nu}\tau_b^{\mu})$, and we used a convention in which  $s^{\mu\nu}=s^{(\mu\nu)}$ for symmetric tensors).} as a consequence of Weyl invariance.

\subsection{Gauge transformations and no-gravitini projection}     
\label{no-gravitini}
A gauge transformation generated by a local, infinitesimal, $su(2,1|2)$-valued zero-form $G$,
\begin{equation}
G=\lambda^a \mathbb{J}_a + \rho^I \mathbb{T}_I + {\overline{\varepsilon}}^i \mathbb{Q}_i - \overline{\mathbb{Q}}^i \varepsilon_i+\lambda \mathbb{Z}\,,
\end{equation}
induces transformations on the component fields $\omega^a$, $A^I$, $e^a$, $\psi_i$ and $b$. While the transformation laws for $\omega^a$, $A^I$ and $b$ are straightforward to read off from the usual rule $\delta \mathbb{A} =d \mathbb{A}+ [\mathbb{A},G]\equiv DG$, extra care is required when handling $e^a$ and $\psi_i$, since this expression only determines the variation of the product $\slashed{e}\psi_i$. In order to see the form of the transformations on the fields $\psi_i$ and $e^a$, we follow the prescription in \cite{Alv14a}, which basically ensures that the vielbein remain invariant under gauge and supersymmetry transformations, but rotate as vectors under the Lorentz subgroup. This is in line with the standard assumption that the metric is unaffected by internal gauge transformations like $U(1)$ and $SU(N)$.

\textbf{Internal gauge symmetry.} The $u(1)$ transformations generated by $\lambda \mathbb{Z}$ affect only the $u(1)$ field $b$, which changes as $\delta b =d \lambda$. Under $su(2)$, the nonzero transformations are
\begin{align}
\delta A^I & = D \rho^I\,, \\
\delta \psi_i & = \frac{i}{2} \rho^I (\sigma_I )_i^{\ j} \psi_j\,, \\
\delta {\overline{\psi}}^i & =- \frac{i}{2}  \overline{\psi}^j\rho^I(\sigma_I)_j^{\ i}\,,
\end{align}
where we have defined the $su(2)$ covariant derivative in the adjoint representation, $D S^I \equiv d S^I + \epsilon^{IJK} A_J S_K$. It is consistent to keep the same notation for the full $so(1,2)\times su(2)$ covariant derivative as long as it is used with the appropriate representation of the argument. In this realization there are no gauginos and the matter field $\psi_i$ transforms in the fundamental of $SU(2)$.

\textbf{Lorentz symmetry.} Under Lorentz rotations, $e^a$ transforms as a vector while the metric $g_{\mu \nu} = \eta_{ab} e^a{}_\mu e^b{}_\nu$ is insensitive to the choice of local orthonormal basis in the tangent space. Consequently, one finds\footnote{In this one-index notation, eq.~(\ref{eq:deL}) is equivalent to $\delta e^a = -\lambda^a{}_b e^b$, with $\lambda^{ab} = -\epsilon^{abc} \lambda_c$.}
\begin{align}
\delta \omega^a & = D \lambda^a \,,\\
\delta e^a & = -\epsilon^{abc} \lambda_b e_c\,,  \label{eq:deL} \\
\delta \psi_i & =- \frac{1}{2} \lambda^a \gamma_a \psi_i\,, \\
\delta {\overline{\psi}}^i & = \frac{1}{2} \lambda^a {\overline{\psi}}^i \gamma_a\,,
\end{align}
where $D \lambda^a \equiv d \lambda^a + \epsilon^{abc} \omega_b \lambda_c$ defines the Lorentz covariant derivative of $\lambda^a$. Note that the matter field $\psi_i$ transforms in the spin-1/2 representation automatically, and not in the adjoint of $so(1,2)$. Obviously, $A^I$ remains invariant under local Lorentz transformations.

\textbf{Supersymmetric rotations.}  Under supersymmetry transformations $\omega^a$, $A^I$ and $b$ change by
\begin{align}
 \delta \omega^{a} & =e^a(\overline{\psi}\varepsilon+ \overline{\varepsilon}\psi)-\epsilon^a{}_{bc}e^b(\overline{\psi}\gamma^c\varepsilon- \overline{\varepsilon}\gamma^c\psi)\,, \\ 
 \delta A^{I} & =-i e^a (\overline{\varepsilon}\sigma^{I}\gamma_a\psi+\overline{\psi}\sigma^I \gamma_a \varepsilon)\,, \\
 \delta b & = - i e^a (\overline{\varepsilon}\gamma_a\psi +\overline{\psi}\gamma_a\varepsilon)\,,
\end{align}
where the $su(2)$ indexes are traced over (and omitted).


A spinor with an extra Lorentz (or spacetime) index, such as $\xi^\alpha_\mu = \gamma_\mu \psi^\alpha$, belongs to the reducible representation $1\otimes 1/2= 1/2 \oplus 3/2$ of the Lorentz group. Hence, it can be uniquely decomposed into its irreducible projections as $\xi_\mu = \Psi_\mu + \Phi_\mu$, where $\Psi_\mu = (P_{1/2})_\mu{}^\nu \xi_\nu$ carries spin-1/2, while  $\Phi_\mu= (P_{3/2})_\mu{}^\nu \xi_\nu$ is the spin-3/2 part (gravitino), with 
\begin{equation}
(P_{3/2})_\mu{}^\nu = \delta^\nu_\mu - \frac{1}{3} \gamma_\mu \gamma^\nu = \delta^\nu_\mu - (P_{1/2})_\mu{}^\nu. \label{projector}
\end{equation}
In our case, $\xi_\mu = e^a{}_\mu \gamma_a \psi = \gamma_\mu \psi$ and therefore the gravitino contribution vanishes identically, $\Phi_\mu \equiv 0$. Thus, a supersymmetry transformation of $\Psi_\mu = e^a{}_\mu \gamma_a \psi $ gives
\begin{equation}
\delta \Psi_\mu = \gamma_a \delta e^a{}_\mu \psi + \gamma_\mu \delta \psi = D_\mu \varepsilon.  \label{projection}
\end{equation}
Finally, multiplying (\ref{projection}) by $P_{1/2}$ yields
\begin{align}
\delta \psi_i &=\frac{1}{3} \gamma^\mu (D_\mu \varepsilon)_i\,,\\
\delta {\overline{\psi}}^i &=-\frac{1}{3}(\overline{\varepsilon}\overleftarrow{D}_\mu)^i \gamma^\mu \,,\\
\delta e^a &= 0\,.
\end{align}
Here $\delta e^a = 0$ ensures that fields remain in a linear representation of SUSY and guarantees that the spin-$3/2$ excitation is not switched on by supersymmetry (no-gravitini condition).  Multiplying (\ref{projection}) by the $P_{3/2}$ projector yields the consistency condition
\begin{equation}\label{cond1}
(P_{3/2})_{\mu}{}^\nu D_\nu \varepsilon = 0.
\end{equation}
This last condition guarantees that the symmetry transformations close off-shell without the need to introduce auxiliary fields. In other words, the fields form a basis for a representation of the $SU(2,1|2)$ gauge group.

Equation \eqref{cond1} can be solved by demanding that $D_\mu \varepsilon$ lives in the kernel of $P_{3/2}$, that is
\begin{equation}\label{cond2}
 D_\mu \varepsilon =\gamma_\mu \chi\,.
\end{equation}
As we will discuss in Section \ref{SUSY-int-cond}, in order to satisfy the integrability condition of the last equation we may demand to make $\varepsilon$ global.

The model has a transparent structure. The presence of fermions requires the introduction of a soldering form (the vielbein) in order to project properties of the dynamical fields in the tangent space onto the base manifold. In particular, the fact that fermions belong to a spin-1/2 representation of the Lorentz group is a feature defined on the tangent space, and the fact that $\mathbb{A}(x)$ is a one-form is a property on the base manifold. Hence, the introduction of vielbeins in the spacetime manifold is required by the presence of fermionic matter, so that a theory that includes fermions necessarily also includes a metric structure, as noted long ago by H. Weyl \cite{Weyl50}.

One of the essential features of gauge theories is the background independence of the gauge symmetry. A gauge transformation
\begin{equation}
\mathbb{A}(x) \rightarrow \mathbb{A}'(x) = g^{-1}(x) \left[ \mathbb{A}(x) +  d  \right] g(x) ,  \label{A-to-A'}
\end{equation}
is the same in any spacetime geometry, i.e., it does not depend on the metric or affine properties of the background.
The invariance of the Yang--Mills Lagrangian,
$\mathcal{L}_\text{YM}=(-1/4)\sqrt{-g}g^{\mu \alpha} g^{\nu \beta}\mbox{Tr} \left( \mathbb{F}_{\mu \nu} \mathbb{F}_{\alpha \beta} \right)$
under (\ref{A-to-A'}) holds at any spacetime point, irrespective of the coordinates, the metric, the background curvature, torsion, etc.
The decoupling between the internal gauge symmetry and the spacetime geometry
is also reflected in the fact that the metric itself is invariant under internal gauge transformations.

Gauge invariance of the metric is ensured if the vielbein is also gauge invariant, $e^a{}_ \mu (x) \rightarrow e'^a{}_\mu (x) = e^a{}_\mu (x)$. Actually, a weaker condition like $e'^a{}_\mu (x) = \Lambda^a{}_b (x) e^b{}_\mu (x)$, with $\Lambda \in SO(1,2)$, would suffice to render
the metric invariant. This is indeed the case when the gauge group includes Lorentz transformations, something that is not often assumed because the Lorentz group is not usually viewed as an internal symmetry. The distinction between internal and spacetime symmetries for the case of the Lorentz group, however, is rather semantic.

The decoupling between the gauge transformations (internal symmetries) and the geometric properties of spacetime (external symmetries) guarantees that gauge invariance holds irrespective of the ``environment.'' Mathematically this is reflected in the fact that a fiber bundle is locally a direct product of a vector space and a manifold: the fibers in the bundle are identical copies of the same algebra, regardless of what the base manifold could be. In the same spirit, here we assume the metric structure to be decoupled from supersymmetry. The vielbein plays the role of a dictionary to translate between the manifold and the tangent space that is not transformed under SUSY. As shown in \cite{Alv11}, this corresponds to a projection of the local SUSY algebra on the spin-1/2 subspace, instead of projecting on the spin-3/2 space, as it is usually done in supergravity \cite{Van81}. The tangent space description in the Riemann-Cartan geometries allows to establish a clear distinction between the vacuum energy $\Lambda$ and a effective cosmological constant $\Lambda_\text{eff}$. In the next section we describe classical solutions that have vanishing vacuum energy but a nonzero $\Lambda_\text{eff}$.

\subsection{Transgression action and charges}   
In order to define conserved charges in the presence of nontrivial boundary conditions it is necessary to add boundary terms to the action to make sure that the action remains stationary on the classical orbits. For trivial conditions, in which all the fields are fixed at the boundary and the space is asymptotically flat, it is often unnecessary to take these precautions, but for asymptotically AdS spacetimes boundary terms are often required. Let us assume a three-dimensional manifold whose local geometry is of the form $\mathcal{M} =\mathbb{R} \times\Sigma$, where $\mathbb{R}$ represents the time direction and $\Sigma$ is the two-dimensional spatial section.  As shown in \cite{Iza05,Iza06a,Mora:2006ka}, the regularized Chern-Simons action is given by the \textit{transgression form} $\mathcal{T}$ defined by
\begin{equation}
\mathcal{T}=L_\text{CS}(\mathbb{A}) +B(\mathbb{A},\overline{\mathbb{A}})- L_\text{CS}(\overline{\mathbb{A}})\,, \label{s1}
\end{equation}
where $\overline{\mathbb{A}}$ corresponds to a fixed classical solution, that matches $\mathbb{A}$ at the boundary and has zero curvature $\mathcal{F}_{\overline{\mathbb{A}}}=0$. The term $B$ lives on the boundary and is such that it makes the action gauge invariant. Varying $\mathcal{T}$ with respect to $\mathbb{A}$ (with $\delta{\overline{\mathbb{A}}}=0$) yields
\begin{equation}
\delta \mathcal{T}=\kappa\left\langle \delta \mathbb{A} \mathcal{F}_\mathbb{A}\right\rangle -\frac{\kappa}{2}d\left\langle \mathbb{A} \delta \mathbb{A}\right\rangle +\delta B.  \label{ds1}
\end{equation}
The first term on the right-hand side vanishes on-shell and therefore $B$ must be such that
\begin{equation}
\left[\delta B - \frac{\kappa}{2} \int_{\partial \mathcal{M}}\left\langle \mathbb{A}\delta \mathbb{A}\right\rangle \right]\biggr\rvert_{\mbox{on-shell}} =0 .  \label{ds}
\end{equation}
The boundary condition $\mathbb{A}\big\rvert _{\partial \mathcal{M}}=\overline{\mathbb{A}}$ means that (\ref{ds}) is fulfilled if
\begin{equation}\label{b}
B=-\frac{\kappa}{2} \int_{\partial \mathcal{M}}\left\langle \mathbb{A}\overline{\mathbb{A}}\right\rangle \,.
\end{equation}

The variation of the transgression form around an arbitrary (infinitesimal) configuration $\delta \mathbb{A}$ will be given by
\begin{equation}
\delta \mathcal{T} = \kappa\langle \delta \mathbb{A}\mathcal{F}_\mathbb{A}\rangle + d\Theta .
\end{equation}
where
\begin{equation}
\Theta=\frac{\kappa}{2}\langle \delta \mathbb{A} \left( \mathbb{A} - \overline{\mathbb{A}}\right)\rangle\,.
\end{equation}
Under gauge transformations $\delta \mathbb{A}=D_\mathbb{A}G$, $\delta \overline{\mathbb{A}}=D_{\overline{\mathbb{A}}}G$, the transgression remains invariant off-shell, $\delta_\text{gauge}\mathcal{T}=0$. Consequently, by Noether's theorem, there is a conserved current $\ast\mathcal{J}=-\Theta$ \cite{Mor04}. Demanding the reducibility condition at spatial infinity, 
\begin{equation}\label{reduc}
D\xi \big\rvert _{\partial \Sigma}=0 \, ,
\end{equation}
the conserved charge can be found as \cite{Barnich:2001jy}
\begin{equation}
Q[\xi] =\kappa\left\langle \left( \mathbb{A}-\overline{ \mathbb{A}}\right) \xi \right\rangle\, . \label{Q[xi]}
\end{equation}
This last expression will be used in section \ref{sect-charges} to compute the global charges of nontrivial vacuum solutions.

\subsection{Mass generation}  
Let us now see how $m$ defined in (\ref{mass}) can be treated as an effective mass induced by spontaneous breaking of Weyl invariance. From the field equation $\mathcal{F}^a=0$ it is straightforward to check that $R^a{}_b e^b = 0$ and therefore the torsion two-form must be covariantly constant in the Lorentz connection,
\begin{equation}
DT^a \equiv R^a{}_b e^b = 0\, .
\end{equation}
In $2+1$ dimensions, the condition $DT^a =0$ can be integrated to
\begin{equation}
T^a = \frac{\epsilon}{z} \epsilon^a{}_{bc} e^b e^c\, ,  \label{T}
\end{equation}
where $z>0$ is an arbitrary integration constant with dimensions of length and $\epsilon = \pm 1$. The introduction of a dimensionful constant fixes the scale for the classical configuration, breaks Weyl invariance and give us an effective mass term for the fermion
\begin{equation}
 m=-\frac{3\epsilon}{2z}\,.
\end{equation}

As will seen below, configurations with non-trivial torsion also contribute to the effective cosmological constant, i.e. the Riemannian curvature. A pure gravity model that exhibits a similar feature is the Mielke-Baekler model \cite{Mielke:1991nn}, see \cite{Blagojevic:2003vn} for the CS formulation of it. In section \ref{solsvacuum} we will comment on the suitable curved background geometries to do perturbation theory. In order to have a well-defined variational principle respecting gauge invariance, the CS terms in a non-compact manifold must be supplemented with a boundary term and appropriate boundary conditions that regularize the action \cite{Mor04}.


\section{Vacuum solutions}     
\label{sec:vacsol}
The field equations (\ref{eom1}) in the matter-free sector, $\psi=0=\overline{\psi}$, imply that spacetime is locally Lorentz flat ($R^{ab}=0$) and  $SU(2)$ flat ($F^I=0$). The interesting point is, however, that this does not necessarily imply a trivial geometry or a trivial $SU(2)$ configuration. The field equations admit other nontrivial solutions depending on the topology and boundary conditions. Moreover, these equations do not completely determine the metric structure and there is a large family of nontrivial solutions that solve them, as discussed in this section.

\subsection{Lorentz-flat geometry} \label{solsvacuum}     
As shown in \cite{Alv14a}, the most general $2+1$ geometry compatible with $R^{ab}=0$ is a geometry of constant negative (Riemann) curvature, i.e., AdS$_3$. Minkowski space is also allowed as a limiting case of vanishing cosmological constant. This can be seen as follows.

The Lorentz connection $\omega$ can be uniquely split into a torsion-free part and the \emph{contorsion} as
\begin{equation}
\nwse{\omega}{a}{b} = \nwse{\mathring{\omega}}{a}{b} + \nwse{\kappa}{a}{b},
\label{eq:cont}
\end{equation}
where $\mathring{\omega}$ is defined by $de^a+\nwse{\mathring{\omega}}{a}{b} e^b \equiv 0$, and the torsion 2-form is $T^a= \nwse{\kappa}{a}{b} e^b$. The torsion-free condition can be solved for $\mathring{\omega}$ in terms of the vielbein and a symmetric affine connection (Christoffel symbol). The Lorentz curvature $R^{ab}$ splits into the torsion-free (Riemannian) curvature $\mathring{R}^{ab}$ and torsion-dependent terms,
\begin{equation}
R^{ab}=\mathring{R}^{ab} + \mathring{D}\kappa^{ab} + \nwse{\kappa}{a}{c} \kappa^{cb}.\label{R}
\end{equation}
Clearly the Lorentz-flat condition $R^{ab}=0$ does not necessarily imply $\mathring{R}^{ab}=0$. The Lorentz-flat condition (\ref{R}), however, implies that the torsion-free connection $\mathring{\omega}^{ab}$ is not generically flat, but $\mathring{R}^{ab} = -\mathring{D} \kappa^{ab} - \nwse{\kappa}{a}{c} \kappa^{cb}$. From (\ref{T}), the contorsion can be written as
\begin{equation}
\kappa^a{}_b=-\frac{\epsilon}{l} \epsilon^a{}_{bc} e^c, \label{kappa}
\end{equation}
where we reserved the constant $l$ for the vacuum case and $l>0$. It can be directly checked that $\mathring{D} \kappa^{ab}=0$ and finally
\begin{equation}
\mathring{R}^{ab} = -\frac{1}{l^2} e^a e^b \, .\label{R=ee}
\end{equation}
The torsion-free part of the Lorentz connection defines the Riemann tensor that accounts for the purely metric (torsion free) curvature,
\begin{equation}
\mathcal{R}^{\alpha \beta}{}_{\mu \nu} = E^\alpha{}_ a E^\beta{}_b\mathring{R}^{ab}{}_{\mu \nu} .  \label{Riemann}
\end{equation}
Combining (\ref{R=ee}) and (\ref{Riemann}), the Riemann tensor for a Lorentz-flat connection is found to be \cite{Alv14a}
\begin{equation}
\mathcal{R}^{\alpha \beta}{}_{\mu \nu}=-\frac{1}{l^2}\left( \delta^\alpha _\mu \delta^\beta_\nu - \delta^\alpha _\nu \delta^\beta_\mu \right).
\end{equation}
Therefore, even if the contribution to the vacuum energy from the fermion condensate were to vanish ($\overline{\psi}\psi=0$), there is an  effective cosmological constant $\Lambda_\text{eff}=-\frac{1}{l^2}$, where $l$ is an arbitrary integration constant. The solution with flat Riemann curvature can also be accommodated by taking $\epsilon=0$ (or $l\rightarrow \infty$). Note that, while there is a sign ambiguity in the torsion ($\epsilon=\pm1$), no such ambiguity exists for the curvature, which means that this result is not true for $\Lambda>0$: de~Sitter spacetime is not a Lorentz-flat geometry.

Considering that the symmetry used to define the model is a superextension of Lorentz symmetry, it is interesting that either flat or negative curvature spaces could emerge spontaneously. Positive curvature, however, is not allowed. We can compare this fact with the four-dimensional case in which de Sitter is not favored by supersymmetry either \cite{Chamseddine:2013hwa}.

Conversely, (\ref{R=ee}) implies that any simply connected patch of three-dimensional anti-de~Sitter space can be endowed with a flat Lorentz connection, just like any patch of Minkowski space. This result can be seen as the Lorentzian version of Adams' theorem, which states that $S^3$ is parallelizable, i.e., it can be endowed with a globally defined flat $SO(3)$ connection \cite{Ada58,Ada60}. This theorem is only valid for $S^0, S^1, S^3$ and $S^7$, so it should not surprise us to have also a similar conclusion in $D=7$ and in no other cases.

In the presence of matter the fermion condensate $\left\langle \overline{\psi}\psi\right\rangle$ relates to the curvature of space and the magnitude of the torsion by means of,
\begin{equation}
\mathring{R}^{ab} =( 2\left\langle \overline{\psi}\psi\right\rangle -\frac{1}{z^2}) e^a e^b \, ,\label{oR=ee}
\end{equation}
and therefore, the effective cosmological constant is
\begin{equation}
 \Lambda_\text{eff}=2\left\langle \overline{\psi}\psi\right\rangle - \frac{1}{z^2} \,.
\end{equation}
This implies that, in order to avoid the appearance of a tachyonic mass term, the following condition has to be satisfied,
\begin{equation}
\frac{1}{z^2} = 2\left\langle \overline{\psi}\psi\right\rangle - \Lambda_\text{eff} \ge 0\, ,
\end{equation}
or $\Lambda_\text{eff}\leq 2\left\langle \overline{\psi}\psi\right\rangle$. There are three different cases: First, if $\left\langle \overline{\psi}\psi\right\rangle < 0$ only AdS spaces are allowed. Second, if $\left\langle \overline{\psi}\psi\right\rangle=0$, then spacetimes with $\Lambda_\text{eff}\leq 0$ are allowed. Finally, for $\left\langle \overline{\psi}\psi\right\rangle > 0$, $\Lambda_\text{eff}$ can take any values in the range $-\infty < \Lambda_\text{eff}\leq 2\left\langle \overline{\psi}\psi\right\rangle$ allowing for metrics that include flat, negative, and a small window of positive curvature spacetimes. 

Summarizing this section, the general solution for the matter-free equations is a spacetime that is locally AdS$_3$ (or Minkowski), where the cosmological constant $\Lambda = -1/l^2$ is an arbitrary integration parameter. This family of geometries includes AdS$_3$ with or without identifications, in particular the 2+1 black hole \cite{Banh93b} and spinning point particles \cite{Mi09}. Since the starting point is only an extension of Lorentz symmetry compatible with either flat, dS or AdS spaces, we find it interesting that the vacuum admits a dynamically selected curved background and rules out de Sitter spacetime.

\subsection{The $2+1$ black hole as a Lorentz-flat geometry}    
The 2+1 black hole is locally AdS$_3$, and the local frame that corresponds to the rotating solution  reads \cite{Ban92}
\begin{align}
e^0&=fdt\,,\\
e^1&=f^{-1}dr\,,\\
e^2&=r\left( d\varphi +N^{\varphi} dt\right),
\end{align}
where
\begin{align}
f(r) & = \left( \frac{r^2}{l^2}-M+\frac{J^2}{4r^2}\right)^{1/2},\\
N^{\varphi} & = -\frac{J}{2r^{2}},
\end{align}
and $(M,J)$ are integration constants corresponding to the mass and angular momentum. The vanishing torsion condition, $de^a+\mathring{\omega }^a{}_b e^{b}=0$, can be solved for the connection yielding
\begin{align}
\mathring{\omega}^0{}_1 &=\frac{r}{l^2}dt-\frac{J}{2r}d\varphi \,,\\
\mathring{\omega}^1{}_2 &=-f d\varphi \,,\\
\mathring{\omega }^2{}_0&=  -\frac{J }{2f r^2}dr\,.  \label{omega-bar}
\end{align}
The corresponding Riemannian two-form has constant, negative (or zero) curvature, $\mathring{R}^{ab}=-l^{-2}e^ae^{b}$. On the other hand, the full Lorentz connection $\omega^a{}_b$, including the contorsion (\ref{kappa}), reads
\begin{align}
\omega^0{}_1& =\left(\frac{r}{l} -\epsilon \frac{J}{2r} \right) \left[\frac{1}{l}dt +\epsilon d\varphi \right]\,,\label{omega0}\\
\omega^1{}_2 & = -f\left[ \frac{\epsilon }{l}dt+d\varphi \right]\,,\label{omega1}\\
\omega^0{}_2& = -\frac{1}{l f}\left( \frac{J l}{2r^2}+ \epsilon \right) dr  \,,\label{omega2}
\end{align}
and is explicitly checked to be flat, $R^{ab}=0$.

Other black holes solutions in the presence of torsion have also been found in the Mielke-Baeckler model \cite{Garcia:2003nm,Mielke:2003xx}.

\subsection{Flat $su(2)$ sector} \label{su(2)solution}    
In addition to being locally AdS$_3$, the vacuum solutions have a locally flat $SU(2)$ connection. This connection is locally pure gauge and therefore can be gauged away in any simply connected patch. But the possibility of gauging it away everywhere depends on the topology of the manifold.

Since $su(2)$ and $so(1,2)$ are locally isomorphic, and the corresponding generators $1/2\sigma_A$ and  $1/2 \gamma_a$ are the same up to factors of $\pm i$ [cf. eq. (\ref{J-T})], one can use the connection (\ref{omega0})--(\ref{omega2}) to tailor the $su(2)$ field $A^I$ as
\begin{align}
A^1&=-i\frac{\eta }{hs}\left( 1-\eta s V^\varphi\right) dr\,,\label{A1}\\
A^2&=-h\left[ \frac{\eta }{s }dt+d\varphi \right]\,,\label{A2}\\
A^3&=-i\frac{\eta r}{s}\left(1+\eta sl V^\varphi\right) \left[ \frac{\eta }{s}dt+d\varphi \right]\,,  \label{A3}
\end{align}
where $\eta =\pm 1$, $s$ is an arbitrary length scale (not necessarily equal to $l$), and
\begin{align}
h(r) & = \left( \frac{r^{2}}{s^2}-W+\frac{K^2}{4r^2} \right)^{1/2}, \\
V^{\varphi} & = - \frac{K}{2r^2}.
\end{align}

The flat $su(2)$ solution \ref{A1}--\ref{A3} makes the asymptotic behavior of the field as
\begin{align}
{\overline{A}}^1&=-i\frac{\eta }{r} dr\,,\label{A1_asymptotic}\\
{\overline{A}}^2&=-\frac{r}{s}\left[\frac{\eta }{s}dt+d\varphi \right]\,,\label{A2_asymptotic}\\
{\overline{A}}^3&=-i\frac{r}{s}\eta\left[\frac{\eta }{s}dt+d\varphi \right].\label{A3_asymptotic}  
\end{align}
Here $(W,K)$ are integration constants. This configuration for $A^I$ closely mimics the Lorentz connection $\omega ^a{}_b$ [cf.~eqs.~(\ref{omega0})--(\ref{omega2})], but the field equations allow nonetheless for independent integration constants ($W,K,s$). As in the Lorentz connection, there is a sign ambiguity ($\eta $) in the solution for $F^I=0$, but in this case it is not related to another structure because in $SU(2)$ there is no analogue for the local frame or the torsion. Additionally, the solution \eqref{A1}--\eqref{A3} allows another sign freedom that corresponds to the choice of sign in the square root to define $h$. There is no analogue of this sign freedom in the Lorentz connection, since it would amount to choosing a local basis in tangent space with the opposite handedness relative to the coordinate basis.

\subsection{Conserved Charges}   
\label{sect-charges}
The nontriviality of the configuration can be assessed by computing the conserved charges (\ref{Q[xi]}). For a generator $\xi=\alpha^a \mathbb{J}_a+\beta^I \mathbb{T}_I \in so(1,2)\oplus su(2)$ we explicitly have
\begin{equation}\label{QNoether}
Q[\xi] = \frac{\kappa}{4} \left[ (\omega^a-\overline{\omega}^a)\alpha_a +(A^I-\overline{A}^I) \beta_I \right].
\end{equation} 
This charge requires the definition of the asymptotic behavior of $\overline{\omega}^a$ and $\overline{A}^{I}$ \cite{Barnich:2001jy}, given by the leading order in $r$ for $r\rightarrow \infty$, where these connections approach those of the massless black hole and the uncharged $su(2)$ solution. Therefore
\begin{align}
\omega^0-\overline{\omega}^0 & =-\frac{Ml}{2r} \left(\frac{\epsilon}{l} dt + d\varphi \right) + O\left(r^{-3}\right) \,, \\
\omega^1-\overline{\omega}^1 & =\frac{l^2}{2r^3}\left(\epsilon M+\frac{J}{l}\right) dr+O(r^{-5}) \,, \\
\omega^2-\overline{\omega}^2 & =-\frac{J}{2r} \left(\frac{\epsilon}{l} dt + d\varphi \right) \,,
\end{align}
and
\begin{align}
A^1-\overline{A}^1 & =-i\frac{s^2}{2r^3}\left(\eta W+\frac{K}{s}\right) dr+O(r^{-5}) \,, \\
A^2-\overline{A}^2 & =\frac{Ws}{2r} \left(\frac{\eta}{s} dt + d\varphi \right) + O\left(r^{-3}\right) \,,\\
A^3-\overline{A}^3& =i\frac{K}{2r} \left(\frac{\eta}{s} dt + d\varphi \right) \,.
\end{align}
Now, the reducibility condition \eqref{reduc} for $\xi$ implies in the asymptotic region
\begin{equation}
d\alpha^a +\epsilon^a{}_{bc}\omega^b \alpha^c =0\,,  \quad d\beta^I + \epsilon^I{}_{JK}A^J \beta^K =0\,.  \label{alpha_beta}
\end{equation}
The asymptotic solutions are
\begin{align}
\alpha^0&=\epsilon c_1\left(r+\frac{\epsilon lJ-l^2M}{2r}\right)+O\left( r^{-2}\right)\,,\\
\alpha^1&=0\,,\\
\alpha^2&=c_1r+O\left( r^{-1}\right)\,,   \label{alpha_sol}
\end{align}
and
\begin{align}
\beta^1&=0\,,\\
\beta^2&=\eta c_2\left(r+\frac{\eta s W-l^2W}{2r}\right)+O\left( r^{-2}\right)\,,\\
\beta^3&=i\eta c_2r+O\left( r^{-1}\right)\,. \label{beta_sol}
\end{align}
with $c_1$ and $c_2$ some arbitrary constants. Finally, the Noether charge (\ref{QNoether}) is found to be
\begin{align}
Q&=\kappa \left(\frac{c_1}{8l} (\epsilon Ml - J) + O\left(r^{-2}\right) \right) \left[ \epsilon dt + l d\varphi \right]\nonumber\\
 &+ \kappa \left(\frac{c_2}{4s}(\eta yW- K) + O\left( r^{-2}\right) \right) [ \eta dt +s d\varphi ] \label{charge}.
\end{align}
This charge must be integrated on a circle at spatial infinity of a time slice, to obtain the conserved quantities associated to the two symmetry groups,
\begin{equation}
\int_{S^1_{\infty}} Q = \frac{\pi\kappa}{2}(c_1 q_{SO(1,2)}  + c_2 q_{SU(2)})\,, \label{Noether}
\end{equation}
where
\begin{align}
q_{SO(1,2)}&=\epsilon Ml - J \,,\\
q_{SU(2)}&= \eta sW- K\,. \label{q-q}
\end{align}

\begin{figure}[thpb]
\centering
\begin{tikzpicture}[
     x=\columnwidth/6,
     y=\columnwidth/6,
     inner sep=2pt,
     outer sep=2pt,
     border rotated/.style = {shape border rotate=180}]
\fill [blueish] (-2,2) -- (2,2) -- (0,0) -- cycle;
\fill [yellowish] (-2,-2) -- (2,-2) -- (0,0) -- cycle;
\draw [<->,thick] (-2.2,0) -- (2.2,0) node [below] {$J/l$};
\draw [<->,thick] (0,-2.2) -- (0,2.2) node [left] {$M$};
\draw [very thick] (-2,-2) -- node [sloped, above] {1ks}(-0.5,-0.5) -- (0,0) -- node [sloped, below] {1ks} (2,2);
\draw [very thick] (-2,2) -- node [sloped, below] {1ks}
  (0,0) -- (0.5,-0.5) -- node [sloped, above] {1ks}  (2,-2);
\draw [very thick,dotted] (1,-2)  -- node [sloped, above] {2ks} (0,-1) -- (-0.5,-0.5);
\draw [very thick,dotted] (-1,-2) -- node [sloped, above] {2ks} (0,-1) -- (0.5,-0.5);
\fill [blueish,above=3] (-0.1,0.28) -- (0.1,0.28) -- (0.01,0.11)-- (-0.01,0.11) -- cycle;
\fill (0,0) circle [radius=0.067] node [above=4] {2ks};
\node [right=2] at (0.5,-0.5) {3ks};
\node [regular polygon, regular polygon sides=3, border rotated, fill=black, inner sep=1.5] at (0.5,-0.5) {};
\node [left=2] at (-0.5,-0.5) {3ks};
\node [regular polygon, regular polygon sides=3, border rotated, fill=black, inner sep=1.5] at (-0.5,-0.5) {};
\fill [yellowish] (-0.1,-0.66) -- (0.1,-0.66) -- (0.01,-0.85)-- (-0.01,-0.85) -- cycle;
\node [above=3] at (0,-1) {4ks};
\node [regular polygon, regular polygon sides=4, fill=black, inner sep=2] at (0,-1) {};
\draw [dashed, very thick] (-1.5,2) -- (2,-1.5);
\draw [dashed, very thick] (-1.5,-2) -- (2,1.5);
\end{tikzpicture}
\caption{Mass-angular momentum phase diagram for three-dimensional solutions of gravity.
The upper wedge (light blue, or darker gray in the printed version), $M>|J|/l$, corresponds to nonextremal black holes configurations. The lower wedge (light yellow in electronic version), $M<-|J|/l$, corresponds to point particles.
Left and right wedges, $|M|<|J|/l$, are unphysical configurations. At $M=-1$ we have anti-de Sitter spacetime (square).
Solid lines correspond to $M\pm |J|/l=0$.
Dotted lines in the lower wedge correspond to $M\pm |J|/l = -1$.
The dashed lines correspond to $-1<M\pm |J|/l<0$.
A generic configuration has no globally-defined Killing spinors (ks), but there may be up to 4 ks for special values of $M$ and $J$. Further explanation in the main text.}
\label{figbhsector}
\end{figure}
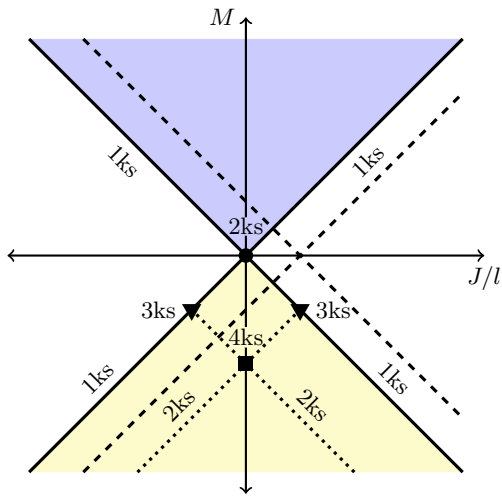

Each of the two symmetry groups have a single Casimir operator and this is reflected in the two charges produced by Noether's procedure. In order to see how these charges determine the configuration, let us consider the charge associated to the Lorentz group, which is determined by two continuous parameters ($M$,$J/l$) and one sign ($\epsilon$). For a fixed value of $q_{SO(1,2)}$, there are two sets of points in the $(M - J/l)$ plane that correspond to it,
\begin{equation}
M = \pm \frac{1}{l} \left( J + q_{SO(1,2)} \right). \label{charges}
\end{equation}
These are two straight lines of slope $\pm 1$ that intersect at the point $(M,J/l)=-(0,q_{SO(1,2)}/l)$.  As shown in Fig.~\ref{figbhsector}, these lines (dashed) correspond to all states for some negative value of $q_{SO(1,2)}$, which include black holes (upper wedge), point particles (lower wedge) and unphysical states (left and right wedges). Each of these lines intersects an extremal black hole, $M=|J|/l$, or an extremal spinning particle, $M=-|J|/l$ \cite{Mi09}.  As will be shown in the next section, for $q_{SO(1,2)}\neq 0$ those extremal states admit-globally defined Killing spinors (BPS states).

A given value of the $SU(2)$ charge also corresponds to two lines in the $(W,K/y)$ plane, but in this case there is no geometric interpretation provided by the metric, which discriminates between black holes, point particles and unphysical states. In contrast with Poincar\'e or AdS gauge theories, here we have only one independent charge asociated to space-time or Lorentz gauge transformations \cite{Blagojevic:2006jk,Ma:2013eaa,Blagojevic:2010jv,Blagojevic:2013aaa}.

It is important to clarify how the flat $SU(2)$ solution should be interpreted. The fact that the fundamental homotopy group of $SU(2)$ is trivial, $\pi_1(SU(2))=0$, tells us that that symmetry is necessarily broken in the solution of Section \ref{su(2)solution}. This is the result of imposing specific asymptotic behavior (\ref{A1_asymptotic})-(\ref{A3_asymptotic}) by demanding $D\beta^I=0$ at spatial infinity. In this sense (\ref{q-q}) is really an $SU(2)$-singlet charge, computed with respect to certain ``orientation'' of the $\beta^I$ parameter. The parameter $\beta^I$ is analogous to the Higgs-like field of the 't Hooft-Polyakov monopole solution \cite{'tHooft:1974qc,Polyakov:1974ek,Abbott:1982jh}. Here we have not included the charge associated with the central generator $\mathbb{Z}$, which could be treated as a $U(1)$ charge in the usual sense and added to the other two charges trivially.

\section{Killing Spinors}        
\label{killingspinors}
If a bosonic system has a classical solution, it is often sufficient to show that it admits globally defined Killing spinors in order to prove perturbative stability. The idea is to embed the theory into a supersymmetric one so that the supersymmetric action is stationary around the classical solution. Then, supersymmetry is typically enough to show that the classical solution is a local energy minimum and therefore perturbatively stable \cite{Witten:1981me}. Using the covariant derivative (\ref{D}), the Killing spinor equation $(D\psi)_i=0$ can be written as\footnote{Recalling that $\gamma$ and $\sigma$ belong to different spaces and therefore act on different indexes of $\psi $, we can safely omit all indexes and simply write $\gamma$ and $\sigma$ both acting on $\psi $ from the left.}
\begin{equation}\label{kseq}
d\psi +\frac{1}{2}\omega^a\gamma_a \psi -\frac{1}{2}A^I \sigma_I \psi =0,
\end{equation}
where $\omega^{a}$ and $A^I$ are given by eqs. (\ref{omega0})--(\ref{A3}).

\subsection{Solutions}    
The general solution for the Killing spinor equation is given by
\begin{equation}\label{ks}
 \psi=U_X U_\gamma U_\sigma U_Y\ \psi_0 ,,
\end{equation}
where $\psi_0$ is a constant spinor and
\begin{align}
U_X&= X\gamma_{-\epsilon} +\frac{1}{X}\gamma_\epsilon,\label{def1}\\
U_Y&=Y\sigma_\eta +\frac{1}{Y}\sigma_{-\eta}, \label{def2}\\
U_\gamma&=\exp \left[ -\theta_{(\epsilon/l)} \gamma_0 \left( \left[ -M+\frac{\epsilon J}{l}\right] \gamma_{-\epsilon} +\gamma_\epsilon \right) \right]\,,\label{def3}\\
U_\sigma&=\exp \left[ -i\theta_{(\eta/s)}\left( \left[ -W+\frac{\eta K}{s} \right]\sigma_\eta +\sigma_{-\eta} \right) \sigma_2 \right]\,.\label{def4}
\end{align}
In (\ref{def1})--(\ref{def4}) we have defined
\begin{align}
X&=\left( f+\frac{r}{l}-\frac{\epsilon J}{2r}\right) ^{1/2}\,,\label{X}\\
Y&=\left( h+\frac{r}{s}-\frac{\eta K}{2r}\right) ^{1/2}\,,\label{Y}\\
\theta_{(v)}&=\frac{1}{2}\left( vt +\varphi \right)\,, \label{theta}\\
\gamma _\epsilon =\frac{1}{2} & \left( 1+\epsilon\gamma_1 \right), \quad \sigma_\eta=\frac{1}{2}\left(1+\eta \sigma_1 \right) .\label{gamma-sigma}
\end{align}
Note that $U_\gamma$ and $U_\sigma$ depend only on $t$ and $\varphi$, while $U_X$ and $U_Y$ depend only on $r$ (see Appendix \ref{app_killing_spinors} for details).

\subsection{Periodicity Conditions}  
\label{periodicity}
Let us examine the periodicity of $U_\gamma$ and $U_\sigma$ under $\varphi \to \varphi +2\pi$ for different values of $(M,J/l)$ and $(W,K/y)$. Under that rotation, the phases of $U_\gamma$ and $U_\sigma$,
\begin{align}
S=-\gamma_0  \left( \left[ -M+\frac{\epsilon J}{l}\right] \gamma_{-\epsilon} +\gamma_\epsilon \right)\, , \\
Z=-i \left( \left[ -W+\frac{\eta K}{s} \right]\sigma_\eta +\sigma_{-\eta} \right) \sigma_2
\end{align}
get multiplied by $2\pi$. There are two possibilities for periodicity to occur: i) if $S^2=-1=Z^2$, in which case the corresponding $U$s would be trigonometric functions of $\theta$, and ii) if $S^2=0=Z^2$, in which case there is no $\varphi$-dependence at all. Direct computation yields
\begin{align}
S^2&= \left[\gamma_0\left(\left[-M+\frac{\epsilon J}{l}\right] \gamma_{-\epsilon} +\gamma_\epsilon \right) \right]^2 \nonumber\\
&=M -\frac{\epsilon J}{l},  \label{cuad1} \\
Z^2&=\left[ i\left( \left[ W-\frac{\eta K}{s}\right] \sigma_\eta -\sigma_{-\eta} \right) \sigma_2 \right]^2 \nonumber\\
&=W-\frac{\eta K}{s}.   \label{cuad2}
\end{align}
In the $(M,J/l)$ plane one can distinguish three different cases: i) $M-\epsilon J/l=-1$, corresponding to two straight lines passing through the AdS point, $(-1,0)$; ii) the generic extremal cases, $M=|J|/l\neq 0$; and iii) the zero mass extremal case, $M=0=|J|$. Three analogous cases can be distinguished in the $(W,K/s)$ plane simply replacing $(\epsilon,l,M,J)$ by $(\eta,s,W,K)$ which, together with the other three, produce nine combined cases. In each of these cases the number of globally-defined Killing spinors is different, as summarized in Table \ref{table-ks}, also depicted in Fig.~\ref{figbhsector} (see Appendix \ref{app_killing_spinors} for details).

In the case $M-\epsilon J/l=-1$, for each value of $\epsilon$ there are two well-defined solutions, the two basis for the constant spinor $\psi_{0}$, represented by the solid lines in Fig.~\ref{figbhsector}. In the second case, $M=|J|/l\neq 0$, there is only one well-defined solution corresponding to the basis spinor in the kernel of $\gamma_\epsilon$, represented by the two lines $M=\pm J/l$. We can also see that at the two black triangles there are three well-defined solutions: two for the value of $\epsilon$ such that $M - J/l = -1$ and one for the value of $\epsilon$ such that $M - J/l = 0$.

Since the $SO(1,2)$ and $SU(2)$ symmetry groups are independent, each one with its own constants of integration, the total number of Killing spinors is just the product of the number of well-defined solutions \eqref{ks} of each sector. The number of complex components of $\psi_0$ is four, two from the spinor index and two from the internal index. The final number of Killing spinors for each case is given in Table~\ref{table-ks}.

\begin{table}[ht]
\begin{center}
\begin{tabular}{|c||c|c|c|c|}
\hline
\diaghead{\theadfont XXXXXXXXX}%
{$(W,K/s)$}{$(M,J/l)$}& $(-1,0)$ & $(0,0)$ & $(\mathbb{R}^+,M)$ \\ \hline\hline
$(-1,0)$ & $16$ & $8$ & $4$ \\ \hline
$(0,0)$ & $8$ & $4$ & $2$ \\ \hline
$(\mathbb{R}^+,W)$ & $4$ & $2$ & $1$ \\ \hline
\end{tabular}%
\end{center}
\caption{{\protect\small {The number of Killing spinors for different values of $(M,J/l)$ and $(W,K/s)$ in the black hole region.}}}
\label{table-ks}
\end{table}

\subsection{SUSY integrability condition}   
\label{SUSY-int-cond}
The number of Killing spinors defines the number of unbroken supersymmetries, i.e., supersymmetries respected by the background. In Section~\ref{no-gravitini} we cast the no-gravitini consistency condition \eqref{cond1} as a kernel equation \eqref{cond2}. By demanding a more restrictive condition, $\chi=c\, \varepsilon$, where $c$ is an arbitrary constant, eq. \eqref{cond2} can be written as
\begin{equation}
d\varepsilon +\frac{1}{2}(\omega ^a -2c\, e^a)\gamma_a\varepsilon -\frac{1}{2}A^I \varepsilon \sigma_A=0\,.
\end{equation}
There are three cases for $c$ of particular simplicity. For $c=0$ this equation obviously reduces to the Killing spinor eq. \eqref{kseq}. The case $c=\epsilon/l$ is similar, in the sense that the equation reduces to the Killing spinor equation for a connection with vanishing Lorentz curvature but with torsion of the opposite sign. The last choice is the intermediate case $c=\epsilon/(2l)$, for which we simply recover the Killing spinor equation in the torsion-free case,
\begin{equation}
d\varepsilon +\frac{1}{2}\mathring{\omega}^a \gamma_a\varepsilon -\frac{1}{2}A^I \varepsilon \sigma_I=0\,.
\end{equation}

These cases are summarized in Table~\ref{cases}, where $\tilde{\omega}^a=\omega^a-2c e^a$, $\tilde{R}^{ab} \equiv d\tilde{\omega}^{ab}+\tilde{\omega}^a{}_c\tilde{\omega}^{cb}$ and $\tilde{T}^a\equiv \tilde{D}e^a$.
\begin{table}
\begin{center}
\begin{tabular}{|c|c|c|c|}
\hline
$c$             & $\tilde{\omega}^a$ & $\tilde{R}^{ab}$ & $\overset{}{\tilde{T}^a}$\\ \hline
0               & $\mathring{\omega}^a+\epsilon/l=\omega^a$ & 0 & $\epsilon/l\ \epsilon^a{}_{bc}e^b e^c$\\ \hline
$\epsilon/(2l)$ & $\mathring{\omega}^a$            & $-1/l^2\ e^a e^b$ & 0\\ \hline
$\epsilon/l$    & $\mathring{\omega}^a-\epsilon/l$ & 0 & $-\epsilon/l\ \epsilon^a{}_{bc}e^b e^c$\\ \hline
\end{tabular}
\end{center}
\caption{Simple cases for the integrability condition.}
\label{cases}
\end{table}

\section{Summary}      
\label{summary} 

We have considered the CS theory for the superalgebra $su( 2,1| 2)$. This algebra can be seen as the minimal supersymmetric extension of the algebra $so(1,2) \oplus su(2)$. This yields an action containing, in addition to the $SU(2)$ connection and $2+1$ gravity, a Dirac field minimally coupled to gravity and to the $SU(2)$ gauge field. As in \cite{Alv11}, the cosmological constant and the mass of the fermion are related and determined by an integration constant, instead of being fundamental parameters in the action.

This system can be viewed as a three-dimensional toy model of a more ``realistic'' four-dimensional theory. However, the Dirac Lagrangian minimally coupled to the $SU(2)$ gauge field also seems appropriate to describe electrons in graphene in the long wavelength limit near the Dirac point, including the possibility of spin-spin coupling mediated by the $SU(2)$ gauge field. This interaction corresponds to assuming the freedom of choosing the spin quantization axis independently at each point in the graphene lattice, as done in the Jordan-Wigner transformation for the Hubbard model \cite{HZ93}. This interaction might produce long range correlations between electron pairs with antiparallel spins in a manner analogous to the Cooper pairs in the BCS theory, in which case, a superconducting phase could exist in graphene at low temperature.

The field equations in the matter-free case obtained by setting to zero the fermions, are those of a locally flat $SU(2)$ connection in a background of locally maximally symmetric three-dimensional spacetime, which includes AdS$_3$, black holes, and point particles (conical singularities) in AdS$_3$, as well as their spinning counterparts. By exploiting the fact that for a particular choice of the torsion a locally AdS$_3$ geometry is Lorentz-flat, a globally nontrivial although locally flat $SU(2)$ connection is constructed mimicking the geometry of a $2+1$ black hole. The solution has a $SU(2)$ charge. However, as discussed in Section \ref{sect-charges}, this is not colored but just an abelian charge from the residual broken symmetry $SU(2)\rightarrow U(1)$. If the black hole is rotating, it is characterized by a combination of the parameters $(M,J)$ and $(W,K)$. For certain specific values of these parameters, the solutions admit globally defined Killing spinors, which means that the corresponding solutions are candidates for perturbatively stable ground state configurations with a number of unbroken supersymmetries.

\begin{acknowledgments}
Discussions with T.Andrade, F.Canfora, G.Giribet, L.Huerta, C.Mart\'{\i}nez, J.Oliva, and M.Valenzuela, are gratefully acknowledged. P.P. is supported by grants from Universidad Andr{\'e}s Bello's Vicerrectorr{\'i}a de Investigaci{\'o}n y Doctorado. P. S-R. is supported by grants from Becas Chile, CONICYT. This work was also partially supported by Fondecyt grant 1140155.  The Centro de Estudios Cient\'{\i}ficos (CECs) is funded by the Chilean Government through the Centers of Excellence Base Financing Program of Conicyt.
\end{acknowledgments}

\appendix

\section{Representation of the $su(2,1|2)$ superalgebra}     
\label{App1}

The following matrices provide a natural representation,
\begin{equation}\label{J-T}
\mathbb{J}_a =\left[ \begin{array}{c|c}
\frac{1}{2} \left( \gamma_a \right)^{\alpha}_{\; \beta} & 0_{2\times 2} \\ \hline
0_{2\times 2} & 0_{2\times 2} \end{array}\right] , \quad \mathbb{T}_I =\left[
\begin{array}{c|c}
0_{2\times 2} & 0_{2\times 2} \\ \hline
0_{2\times 2} & -\frac{i}{2} (u\sigma_I u^\dagger )^{i}_{\ j}\end{array}
\right] ,
\end{equation}
where $\gamma_a$, $a=0,1,2$, are Dirac matrices, with $\alpha,\beta=1,2$, and $\sigma_I$, $I=1,2,3$, are Pauli matrices, with $i,j=1,2$. A metric to raise and lower latin indexes is given by $[u^{ij}]=i\sigma_2$. A generic supermatrix $\mathbb{M}$ has the following index structure:
\begin{equation*}
\mathbb{M} = \left[ \begin{array}{c|c}
M^\alpha{}_\beta & M^\alpha{}_j \\ \hline
M^i{}_\beta & M^i{}_ j%
\end{array}\right] \,.
\end{equation*}
In terms of components, (\ref{J-T}) are given by
\begin{equation}
(\mathbb{J}_a)^A_{\ B}=\frac{1}{2}(\gamma_a)^A_{\ B}\,, \quad (\mathbb{T}_I)^A_{\ B}=-\frac{i}{2}u^{Ai}(\sigma_I)_i{}^j u_{jB}\,,
\end{equation}

A direct calculation shows that\footnote{Flat Lorentz and $SU(2)$ indexes in the adjoint representations are lowered and raised using the Lorentzian and Euclidean metrics $\eta_{ab}$ and $\delta^{IJ}$, respectively.}
\begin{align*}
\left[ \mathbb{J}_a ,\mathbb{J}_b \right] &= \epsilon_{ab}{}^c \mathbb{J}_c\,, \\
\left[ \mathbb{T}_I ,\mathbb{T}_J \right] &= \epsilon_{IJ}{}^K \mathbb{T}_K\,,
\end{align*}
and $\left[ \mathbb{J}_a,\mathbb{T}_I \right] =0$. The fermionic generators
\begin{equation}
(\mathbb{Q}^\alpha_i)^A_{\ B}=\delta^A_i \delta^\alpha_B\,,\quad (\overline{\mathbb{Q}}_\alpha^i)^A_{\ B}=\delta^A_\alpha \delta^i_B\,,
\end{equation}
are defined so that
\begin{equation*}
(\overline{\psi}\mathbb{Q})^A_{\ B} = \left[
\begin{array}{c|c}
0_{2\times 2} & 0_{2\times 2} \\ \hline
\overline{\psi}_{B}^A & 0_{2\times 2}
\end{array}
\right] \,, \quad
(\overline{\mathbb{Q}}\psi)^A_{\ B}=\left[
\begin{array}{c|c}
0_{2\times 2} & \psi^A_B \\ \hline
0_{2\times 2} & 0_{2\times 2}
\end{array}
\right] \,.
\end{equation*}
Direct computation gives
\begin{equation*}
\{ \mathbb{Q}_i^\alpha,\mathbb{Q}_j^\beta\} = 0, \quad
\{ \overline{\mathbb{Q}}_\alpha^i,\overline{\mathbb{Q}}_\beta^j\} = 0 ,
\end{equation*}
and
\begin{equation}
 [\{Q_i^\alpha,\overline{Q}^\beta_j\}]^A_{\phantom{A} C}=\delta_i^j \delta^A_\beta\delta^\alpha_C+\delta^\alpha_\beta\delta^A_i \delta^j_C\,,
\end{equation}
The completeness relations for Dirac and Pauli matrices can be used to recast this as
\begin{equation*}
\{ \mathbb{Q}_i^\alpha,\overline{\mathbb{Q}}_\beta^j\} =\delta_i^j(\gamma ^a)^\alpha{}_\beta\mathbb{J}_a-i\delta_\beta^\alpha (\sigma^I)_i^{\ j}\mathbb{T}_I-i\delta_i^j\delta_\beta^\alpha\mathbb{Z}\,,
\end{equation*}%
where $\mathbb{Z}$ is a new bosonic generator represented by a diagonal matrix with vanishing supertrace,
\begin{equation*}
\mathbb{Z}^A_{\ B}=\frac{i}{2}(\delta^A_\alpha \delta^\alpha_B+\delta^A_i \delta^i_B)\,.
\end{equation*}
This generator is a central charge that commutes with all generators in the superalgebra. The only remaining commutators are
\begin{align*}
 [ \mathbb{J}_a,\mathbb{Q}_i^\alpha] &=-\frac{1}{2}\left(\gamma_a\right)^\alpha_{\ \beta}\mathbb{Q}_i^\beta\,, \quad [ \mathbb{J}_a,\overline{\mathbb{Q}}_\alpha^i] =\frac{1}{2}\left( \gamma_a\right)^\beta_{\ \alpha}\overline{\mathbb{Q}}_\beta^i\,,\\
 [ \mathbb{T}_I,\mathbb{Q}_i^\alpha] &=\frac{i}{2}\left(\sigma_I\right)_i^{\ j}\mathbb{Q}_j^\alpha\,, \quad [\mathbb{T}_I,\overline{\mathbb{Q}}_\alpha^i]=-\frac{i}{2}\left(\sigma_I\right)_j^{\ i}\overline{\mathbb{Q}}_\alpha^j\,.
\end{align*}
This completes the algebra. It can be directly checked that each of these generators have vanishing supertrace.

\section{Killing spinors} \label{app_killing_spinors}     
In this Appendix we present a detailed computation of the Killing spinors mentioned in Table \ref{table-ks}. The radial, time, and angle components of this equation read
\begin{align}
0=&\partial_r \psi + \frac{1}{2}\left( \frac{\epsilon}{l}-N^\varphi \right)f^{-1}\gamma_1 \psi \nonumber\\
&-\frac{1}{2}\left(\frac{\eta }{s}-V^\varphi\right) h^{-1}\sigma _{1}\psi  \,,  \label{eq:KSr} \\
0=&\partial_t \psi +\frac{\epsilon}{2l}\left( f\gamma_0 +r\left( \frac{\epsilon}{l}+ N^\varphi\right)\gamma_2 \right) \psi \nonumber\\
& +\frac{\eta}{2s}\left( ih\sigma_2 -r\left( \frac{\eta}{s} + V^\varphi \right) \sigma_3 \right)  \psi  \,, \label{eq:KSt} \\
0=&\partial_\varphi \psi + \frac{1}{2}\left( f\gamma_0 + r\left( \frac{\epsilon }{l} + N^\varphi \right) \gamma_2 \right) \psi \nonumber\\
&+\frac{1}{2}\left( ih\sigma_2 -r\left( \frac{\eta }{s}+V^\varphi\right) \sigma_3 \right) \psi  \,. \label{eq:KSphi}
\end{align}

With $X$ and $Y$ defined in (\ref{X},\ref{Y}), \eqref{eq:KSr} becomes
\begin{equation}
\partial_r \psi =\frac{d}{dr} \left( -\epsilon \gamma_1 \ln X\right) \psi + \frac{d}{dr}\left( \eta \sigma_1 \ln Y\right) \psi , \label{eq:KSr'}
\end{equation}
and the solution of \eqref{eq:KSr'} can be written as
\begin{equation}
\psi =U_X  U_Y \xi\,.
\label{eq:psiansrf}
\end{equation}
Here $\xi $ is an $r$-independent spinor with $U_X$ and $U_Y$ defined in (\ref{def1}) and (\ref{def2}). Replacing \eqref{eq:psiansrf} in (\ref{eq:KSt}) and (\ref{eq:KSphi}), and using the properties of these projectors \footnote{These projectors satisfy
(i)~$\gamma_{\pm}^2 =\gamma_{\pm }$,
(ii)~$\gamma_{\pm }\gamma_{\mp} = 0$,
(iii)~$\gamma_+ +\gamma _- = \um$,
(iv)~$\gamma_{0,2}\gamma_{\pm} =\gamma _{\mp }\gamma_{0,2}$,
(v)~$\gamma_1 \gamma_{\pm} =\pm \gamma_1$,
and similarly for $\sigma_{\pm} $.}
leads to
\begin{align*}
0=&\partial_t \xi +\frac{\epsilon }{2l}\gamma_0 \left[ \left( -M+\frac{\epsilon J}{l}\right) \gamma_{-\epsilon} +\gamma_{\epsilon} \right] \xi\\
&\quad+ \frac{i\eta}{2s} \sigma_2\left[ \left( -W+\frac{\eta K}{s }\right) \sigma_\eta -\sigma_{-\eta} \right]  \xi\,, \\
0=&\partial_\varphi \xi +\frac{1}{2}\gamma_0 \left[ \left( -M+\frac{\epsilon J}{l}\right) \gamma_{-\epsilon} +\gamma_{\epsilon}\right] \xi\\
&\quad+ \frac{i}{2} \sigma _{2}\left[ \left( W-\frac{\eta K}{s} \right) \sigma_\eta -\sigma_{-\eta }\right] \xi\,,
\end{align*}
whose solution is given by \eqref{ks}. Next we present those solutions with well defined periodicity conditions for different values of $(Ml,J)$ and $(Wl,K)$. As the $SO(1,2)$ and $SU(2)$ sector are decoupled, we will consider in detail only the cases where $M=W$ and $\left\vert J\right\vert l=\left\vert K\right\vert s $. The remaining cases in Table \ref{table-ks} can be obtained from these in a straightforward way.

\subsection{Case $M=W=-1$; $J=K=0$}     
\label{m=q=-1} In this case, the functions $X$ and $Y$ take the form
\begin{eqnarray*}
X &=&\left( \frac{r}{l}+n\right) ^{1/2}\mbox{ \ \ },\mbox{ \ \ }n=\left(\frac{r^{2}}{l^{2}}+1\right) ^{1/2} \\
Y &=&\left( \frac{r}{s }+\tilde{n}\right) ^{1/2}\mbox{ \ \ },\mbox{ \ \ }\tilde{n}=\left( \frac{r^{2}}{s ^{2}}+1\right) ^{1/2},
\end{eqnarray*}
and (\ref{ks}) reduces to
\begin{align}
\psi & =\left[ \left( \frac{n+1}{2}\right) ^{1/2}-\epsilon \left( \frac{n-1}{2}\right) ^{1/2}\gamma _{1}\right]\nonumber\\
&\quad \times\left( \cos \theta_{(\epsilon/l)} -\gamma _{0}\sin \theta_{(\epsilon/l)} \right)\nonumber\\
&\quad \times\left[ \left( \frac{\tilde{n}+1}{2}\right)^{1/2}+\eta \left( \frac{\tilde{n}-1}{2}\right) ^{1/2}\sigma _{1}\right]  \nonumber\\
&\quad \times\left( \cos \theta_{(\eta/s)} -i\sigma _{2}\sin \theta_{(\eta/s)} \right)  \psi _{0}\,.
\end{align}
As $\psi _{0}$ has the form
\begin{equation*}
\psi _{0}=\left(
\begin{array}{c}
a \\
b%
\end{array}%
\right) \otimes \left(
\begin{array}{c}
c \\
d%
\end{array}%
\right) \,,
\end{equation*}
with $a,b,c,d$ arbitrary real numbers, it can be spanned in a four dimensional basis. Therefore, there are four spinors for each value of $\epsilon $ and $\eta $ leading to a total of sixteen Killing spinors.

\subsection{Case $M=J=W=K=0$} \label{m=q=0}     

In this case $U_\gamma=\exp \left[ -\frac{1}{2}\theta_{(\epsilon/l)}\left( \gamma_0+\epsilon \gamma_2\right) \right]$ and $U_\sigma=\exp \left[ \frac{i}{2}\theta_{(\eta/s)}\left(\sigma _{2}-i\eta \sigma _{3}\right) \right]$. Since $\left( \gamma _0 +\epsilon \gamma_2 \right) $ is nilpotent, we can write $U_\gamma=\um-\frac{1}{2}\theta_{(\epsilon/l)}\left( \gamma _{0}+\epsilon \gamma _{2}\right) $, and similarly for $U_\sigma$. Hence, in order to get rid of the linear dependence of $\psi $ in $\theta_{(\epsilon/l)}$ and $\theta_{(\eta/s)}$, $\psi _{0}$ must be in the kernel of $\left( \gamma _{0}+\epsilon \gamma _{2}\right) $ and $\left(\sigma _{2}-i\eta \sigma _{3}\right) $, i.e.,
\begin{equation*}
\left( \gamma_0 +\epsilon \gamma_2 \right) \psi_0 = 0 = \psi_0 \left(\sigma _{2}-i\eta \sigma _{3}\right)\,,
\end{equation*}
which is satisfied provided $\psi_0$ is one of the eigenvector of $\gamma_1$ and $\sigma_1$ depending on $\epsilon$ and $\eta $.
Hence $\psi _{0}$ can have the form
\begin{equation*}
\psi _{0}^{(\epsilon ,\eta )}=\left(
\begin{array}{c} 1 \\
-\epsilon
\end{array}
\right) \otimes \left(
\begin{array}{c}
1 \\
\eta%
\end{array}%
\right) \,,
\end{equation*}
As in this case $X=\left( \frac{2r}{l}\right) ^{1/2}$ and $Y=\left( \frac{2r}{s}\right) ^{1/2}$, we obtain
\begin{equation}
\psi =\frac{2r}{\sqrt{ls}}\psi _0^{(\epsilon ,\eta )}\,.  \label{ks2}
\end{equation}
Therefore, in this case there are four Killing spinors, one for each value of $\epsilon $ and $\eta $.

\subsection{Case $M,W>0$; $M=|J|/l$, $W=|K|/y$}

Let us consider the first the option $M=J/l$, $W=K/y$. Then, \eqref{cuad1} and \eqref{cuad2} take the form
\begin{align}
\left[ \gamma _{0}\left( \left( -M+\frac{\epsilon J}{l}\right) \gamma_{-\epsilon }+\gamma _{\epsilon }\right) \right] ^{2}& =M\left(1-\epsilon \right)\,, \nonumber\\
\left[ -i\sigma _{2}\left( \left( W-\frac{\eta K}{s}\right) \sigma _{\eta }+\sigma_{-\eta }\right) \right] ^{2}& =W\left( 1-\eta \right)\,.
\notag
\end{align}%
Nilpotency is achieved in this case for $\epsilon =\eta =1$ leading to $\psi=U_X^{(+)}U_\gamma^{(+)} U_\sigma^{(+)} U_Y^{(+)}\psi_0$, where $U_X^{(+)}=\left.U_X\right|_{\epsilon=+1}$ and $U_Y^{(+)}=\left.U_Y\right|_{\eta=+1}$. Since $\theta_{(1/l)}\gamma _{0}\gamma_{+}$ and $\theta_{(1/s)}\sigma _{-}\sigma_{2}$ are nilpotent,
\[
U_\gamma^{(+)}=1-\theta_{(1/l)}\gamma_0\gamma_+, \; \mbox{and }\; U_\sigma^{(+)}=1+i \theta_{(1/s)}\sigma_- \sigma_2.
\]
Hence, $\psi _{0}$ must be in the kernel of $\gamma_+$ and $\sigma_-$,
\begin{equation*}
\gamma _{+}\psi _{0}=0=\psi _{0}\sigma _{-}\,,
\end{equation*}%
which is satisified by
\begin{equation*}
\psi _{0}=\left(
\begin{array}{c}
1 \\
-1
\end{array}
\right) \otimes \left(
\begin{array}{c}
1 \\
1
\end{array}
\right) \,.
\end{equation*}%
As in this case $X=\sqrt{\frac{2r}{l}-\frac{Ml}{r}},Y=\sqrt{\frac{2r}{s}-\frac{Ws}{r}}$, we finally arrive to
\begin{equation}
\psi =\sqrt{\left( \frac{2r}{l}-\frac{Ml}{r}\right) \left( \frac{2r}{s}-\frac{Ws}{r}\right) }\psi _0\,.
\end{equation}%
Therefore, in this case there is only one Killing spinor. A similar analysis can be done for all the possible particular cases of $\left\vert J\right\vert =Ml$ and $\left\vert K\right\vert =Ws $ leading essentially to the same result. Hence, the extreme case has always only one well-defined Killing spinor.

\bibliography{biblio}{}

\begin{thebibliography}{56}%
\makeatletter
\providecommand \@ifxundefined [1]{%
 \@ifx{#1\undefined}
}%
\providecommand \@ifnum [1]{%
 \ifnum #1\expandafter \@firstoftwo
 \else \expandafter \@secondoftwo
 \fi
}%
\providecommand \@ifx [1]{%
 \ifx #1\expandafter \@firstoftwo
 \else \expandafter \@secondoftwo
 \fi
}%
\providecommand \natexlab [1]{#1}%
\providecommand \enquote  [1]{``#1''}%
\providecommand \bibnamefont  [1]{#1}%
\providecommand \bibfnamefont [1]{#1}%
\providecommand \citenamefont [1]{#1}%
\providecommand \href@noop [0]{\@secondoftwo}%
\providecommand \href [0]{\begingroup \@sanitize@url \@href}%
\providecommand \@href[1]{\@@startlink{#1}\@@href}%
\providecommand \@@href[1]{\endgroup#1\@@endlink}%
\providecommand \@sanitize@url [0]{\catcode `\\12\catcode `\$12\catcode
  `\&12\catcode `\#12\catcode `\^12\catcode `\_12\catcode `\%12\relax}%
\providecommand \@@startlink[1]{}%
\providecommand \@@endlink[0]{}%
\providecommand \url  [0]{\begingroup\@sanitize@url \@url }%
\providecommand \@url [1]{\endgroup\@href {#1}{\urlprefix }}%
\providecommand \urlprefix  [0]{URL }%
\providecommand \Eprint [0]{\href }%
\providecommand \doibase [0]{http://dx.doi.org/}%
\providecommand \selectlanguage [0]{\@gobble}%
\providecommand \bibinfo  [0]{\@secondoftwo}%
\providecommand \bibfield  [0]{\@secondoftwo}%
\providecommand \translation [1]{[#1]}%
\providecommand \BibitemOpen [0]{}%
\providecommand \bibitemStop [0]{}%
\providecommand \bibitemNoStop [0]{.\EOS\space}%
\providecommand \EOS [0]{\spacefactor3000\relax}%
\providecommand \BibitemShut  [1]{\csname bibitem#1\endcsname}%
\let\auto@bib@innerbib\@empty
\bibitem [{\citenamefont {Martin}(2010)}]{Mar11}%
  \BibitemOpen
  \bibfield  {author} {\bibinfo {author} {\bibfnamefont {S.~P.}\ \bibnamefont
  {Martin}},\ }\href {\doibase 10.1142/9789814307505_0001} {\bibfield
  {journal} {\bibinfo  {journal} {Adv. Ser. Direct. High Energy Phys.}\
  }\textbf {\bibinfo {volume} {21}},\ \bibinfo {pages} {1} (\bibinfo {year}
  {2010})},\ \Eprint {http://arxiv.org/abs/hep-ph/9709356}
  {arXiv:hep-ph/9709356 [hep-ph]} \BibitemShut {NoStop}%
\bibitem [{\citenamefont {Chatrchyan}\ \emph {et~al.}(2013)\citenamefont
  {Chatrchyan} \emph {et~al.}}]{CMS13a}%
  \BibitemOpen
  \bibfield  {author} {\bibinfo {author} {\bibfnamefont {S.}~\bibnamefont
  {Chatrchyan}} \emph {et~al.} (\bibinfo {collaboration} {CMS Collaboration}),\
  }\href {\doibase 10.1007/JHEP03(2013)037, 10.1007/JHEP07(2013)041} {\bibfield
   {journal} {\bibinfo  {journal} {JHEP}\ }\textbf {\bibinfo {volume} {1303}},\
  \bibinfo {pages} {037} (\bibinfo {year} {2013})},\ \Eprint
  {http://arxiv.org/abs/1212.6194} {arXiv:1212.6194 [hep-ex]} \BibitemShut
  {NoStop}%
\bibitem [{\citenamefont {Chatrchyan}\ \emph {et~al.}(2014)\citenamefont
  {Chatrchyan} \emph {et~al.}}]{CMS13b}%
  \BibitemOpen
  \bibfield  {author} {\bibinfo {author} {\bibfnamefont {S.}~\bibnamefont
  {Chatrchyan}} \emph {et~al.} (\bibinfo {collaboration} {CMS Collaboration}),\
  }\href {\doibase 10.1103/PhysRevLett.112.161802} {\bibfield  {journal}
  {\bibinfo  {journal} {Phys. Rev. Lett.}\ }\textbf {\bibinfo {volume} {112}},\
  \bibinfo {pages} {161802} (\bibinfo {year} {2014})},\ \Eprint
  {http://arxiv.org/abs/1312.3310} {arXiv:1312.3310 [hep-ex]} \BibitemShut
  {NoStop}%
\bibitem [{\citenamefont {Aad}\ \emph {et~al.}(2014{\natexlab{a}})\citenamefont
  {Aad} \emph {et~al.}}]{ATL14a}%
  \BibitemOpen
  \bibfield  {author} {\bibinfo {author} {\bibfnamefont {G.}~\bibnamefont
  {Aad}} \emph {et~al.} (\bibinfo {collaboration} {ATLAS Collaboration}),\
  }\href {\doibase 10.1103/PhysRevLett.112.231806} {\bibfield  {journal}
  {\bibinfo  {journal} {Phys. Rev. Lett.}\ }\textbf {\bibinfo {volume} {112}},\
  \bibinfo {pages} {231806} (\bibinfo {year} {2014}{\natexlab{a}})}\BibitemShut
  {NoStop}%
\bibitem [{\citenamefont {Aad}\ \emph {et~al.}(2014{\natexlab{b}})\citenamefont
  {Aad} \emph {et~al.}}]{ATL14b}%
  \BibitemOpen
  \bibfield  {author} {\bibinfo {author} {\bibfnamefont {G.}~\bibnamefont
  {Aad}} \emph {et~al.} (\bibinfo {collaboration} {ATLAS Collaboration}),\
  }\href {\doibase 10.1007/JHEP06(2014)035} {\bibfield  {journal} {\bibinfo
  {journal} {JHEP}\ }\textbf {\bibinfo {volume} {1406}},\ \bibinfo {pages}
  {035} (\bibinfo {year} {2014}{\natexlab{b}})},\ \Eprint
  {http://arxiv.org/abs/1404.2500} {arXiv:1404.2500 [hep-ex]} \BibitemShut
  {NoStop}%
\bibitem [{\citenamefont {Alvarez}\ \emph {et~al.}(2012)\citenamefont
  {Alvarez}, \citenamefont {Valenzuela},\ and\ \citenamefont
  {Zanelli}}]{Alv11}%
  \BibitemOpen
  \bibfield  {author} {\bibinfo {author} {\bibfnamefont {P.~D.}\ \bibnamefont
  {Alvarez}}, \bibinfo {author} {\bibfnamefont {M.}~\bibnamefont {Valenzuela}},
  \ and\ \bibinfo {author} {\bibfnamefont {J.}~\bibnamefont {Zanelli}},\ }\href
  {\doibase 10.1007/JHEP04(2012)058} {\bibfield  {journal} {\bibinfo  {journal}
  {JHEP}\ }\textbf {\bibinfo {volume} {1204}},\ \bibinfo {pages} {058}
  (\bibinfo {year} {2012})},\ \Eprint {http://arxiv.org/abs/1109.3944}
  {arXiv:1109.3944 [hep-th]} \BibitemShut {NoStop}%
\bibitem [{\citenamefont {Alvarez}\ \emph
  {et~al.}(2014{\natexlab{a}})\citenamefont {Alvarez}, \citenamefont {Pais},\
  and\ \citenamefont {Zanelli}}]{Alv13}%
  \BibitemOpen
  \bibfield  {author} {\bibinfo {author} {\bibfnamefont {P.~D.}\ \bibnamefont
  {Alvarez}}, \bibinfo {author} {\bibfnamefont {P.}~\bibnamefont {Pais}}, \
  and\ \bibinfo {author} {\bibfnamefont {J.}~\bibnamefont {Zanelli}},\ }\href
  {\doibase 10.1016/j.physletb.2014.06.031} {\bibfield  {journal} {\bibinfo
  {journal} {Phys. Lett.}\ }\textbf {\bibinfo {volume} {B735}},\ \bibinfo
  {pages} {314 } (\bibinfo {year} {2014}{\natexlab{a}})},\ \Eprint
  {http://arxiv.org/abs/1306.1247} {arXiv:1306.1247 [hep-th]} \BibitemShut
  {NoStop}%
\bibitem [{\citenamefont {Deser}\ \emph
  {et~al.}(1982{\natexlab{a}})\citenamefont {Deser}, \citenamefont {Jackiw},\
  and\ \citenamefont {Templeton}}]{Des82a}%
  \BibitemOpen
  \bibfield  {author} {\bibinfo {author} {\bibfnamefont {S.}~\bibnamefont
  {Deser}}, \bibinfo {author} {\bibfnamefont {R.}~\bibnamefont {Jackiw}}, \
  and\ \bibinfo {author} {\bibfnamefont {S.}~\bibnamefont {Templeton}},\ }\href
  {\doibase 10.1103/PhysRevLett.48.975} {\bibfield  {journal} {\bibinfo
  {journal} {Phys. Rev. Lett.}\ }\textbf {\bibinfo {volume} {48}},\ \bibinfo
  {pages} {975} (\bibinfo {year} {1982}{\natexlab{a}})}\BibitemShut {NoStop}%
\bibitem [{\citenamefont {Deser}\ \emph
  {et~al.}(1982{\natexlab{b}})\citenamefont {Deser}, \citenamefont {Jackiw},\
  and\ \citenamefont {Templeton}}]{Des82b}%
  \BibitemOpen
  \bibfield  {author} {\bibinfo {author} {\bibfnamefont {S.}~\bibnamefont
  {Deser}}, \bibinfo {author} {\bibfnamefont {R.}~\bibnamefont {Jackiw}}, \
  and\ \bibinfo {author} {\bibfnamefont {S.}~\bibnamefont {Templeton}},\ }\href
  {\doibase 10.1016/0003-4916(82)90164-6} {\bibfield  {journal} {\bibinfo
  {journal} {Annals Phys.}\ }\textbf {\bibinfo {volume} {140}},\ \bibinfo
  {pages} {372} (\bibinfo {year} {1982}{\natexlab{b}})}\BibitemShut {NoStop}%
\bibitem [{\citenamefont {Achucarro}\ and\ \citenamefont
  {Townsend}(1986)}]{Ach86}%
  \BibitemOpen
  \bibfield  {author} {\bibinfo {author} {\bibfnamefont {A.}~\bibnamefont
  {Achucarro}}\ and\ \bibinfo {author} {\bibfnamefont {P.}~\bibnamefont
  {Townsend}},\ }\href {\doibase 10.1016/0370-2693(86)90140-1} {\bibfield
  {journal} {\bibinfo  {journal} {Phys. Lett.}\ }\textbf {\bibinfo {volume}
  {B180}},\ \bibinfo {pages} {89} (\bibinfo {year} {1986})}\BibitemShut
  {NoStop}%
\bibitem [{\citenamefont {Achucarro}\ and\ \citenamefont
  {Townsend}(1989)}]{Ach89}%
  \BibitemOpen
  \bibfield  {author} {\bibinfo {author} {\bibfnamefont {A.}~\bibnamefont
  {Achucarro}}\ and\ \bibinfo {author} {\bibfnamefont {P.}~\bibnamefont
  {Townsend}},\ }\href {\doibase 10.1016/0370-2693(89)90423-1} {\bibfield
  {journal} {\bibinfo  {journal} {Phys. Lett.}\ }\textbf {\bibinfo {volume}
  {B229}},\ \bibinfo {pages} {383} (\bibinfo {year} {1989})}\BibitemShut
  {NoStop}%
\bibitem [{\citenamefont {Chamseddine}(1990)}]{Cha90}%
  \BibitemOpen
  \bibfield  {author} {\bibinfo {author} {\bibfnamefont {A.~H.}\ \bibnamefont
  {Chamseddine}},\ }\href {\doibase 10.1016/0550-3213(90)90245-9} {\bibfield
  {journal} {\bibinfo  {journal} {Nucl. Phys.}\ }\textbf {\bibinfo {volume}
  {B346}},\ \bibinfo {pages} {213} (\bibinfo {year} {1990})}\BibitemShut
  {NoStop}%
\bibitem [{\citenamefont {Bañados}\ \emph {et~al.}(1996)\citenamefont
  {Bañados}, \citenamefont {Troncoso},\ and\ \citenamefont {Zanelli}}]{Ba96}%
  \BibitemOpen
  \bibfield  {author} {\bibinfo {author} {\bibfnamefont {M.}~\bibnamefont
  {Bañados}}, \bibinfo {author} {\bibfnamefont {R.}~\bibnamefont {Troncoso}},
  \ and\ \bibinfo {author} {\bibfnamefont {J.}~\bibnamefont {Zanelli}},\ }\href
  {\doibase 10.1103/PhysRevD.54.2605} {\bibfield  {journal} {\bibinfo
  {journal} {Phys. Rev.}\ }\textbf {\bibinfo {volume} {D54}},\ \bibinfo {pages}
  {2605} (\bibinfo {year} {1996})},\ \Eprint
  {http://arxiv.org/abs/gr-qc/9601003} {arXiv:gr-qc/9601003 [gr-qc]}
  \BibitemShut {NoStop}%
\bibitem [{\citenamefont {Troncoso}\ and\ \citenamefont
  {Zanelli}(1998)}]{Tro97}%
  \BibitemOpen
  \bibfield  {author} {\bibinfo {author} {\bibfnamefont {R.}~\bibnamefont
  {Troncoso}}\ and\ \bibinfo {author} {\bibfnamefont {J.}~\bibnamefont
  {Zanelli}},\ }\href {\doibase 10.1103/PhysRevD.58.101703} {\bibfield
  {journal} {\bibinfo  {journal} {Phys. Rev.}\ }\textbf {\bibinfo {volume}
  {D58}},\ \bibinfo {pages} {101703} (\bibinfo {year} {1998})},\ \Eprint
  {http://arxiv.org/abs/hep-th/9710180} {arXiv:hep-th/9710180 [hep-th]}
  \BibitemShut {NoStop}%
\bibitem [{\citenamefont {Troncoso}\ and\ \citenamefont
  {Zanelli}(1999)}]{Tro98}%
  \BibitemOpen
  \bibfield  {author} {\bibinfo {author} {\bibfnamefont {R.}~\bibnamefont
  {Troncoso}}\ and\ \bibinfo {author} {\bibfnamefont {J.}~\bibnamefont
  {Zanelli}},\ }\href {\doibase 10.1023/A:1026614631617} {\bibfield  {journal}
  {\bibinfo  {journal} {Int. J. Theor. Phys.}\ }\textbf {\bibinfo {volume}
  {38}},\ \bibinfo {pages} {1181} (\bibinfo {year} {1999})},\ \Eprint
  {http://arxiv.org/abs/hep-th/9807029} {arXiv:hep-th/9807029 [hep-th]}
  \BibitemShut {NoStop}%
\bibitem [{\citenamefont {Van~Nieuwenhuizen}(1981)}]{Van81}%
  \BibitemOpen
  \bibfield  {author} {\bibinfo {author} {\bibfnamefont {P.}~\bibnamefont
  {Van~Nieuwenhuizen}},\ }\href {\doibase 10.1016/0370-1573(81)90157-5}
  {\bibfield  {journal} {\bibinfo  {journal} {Phys. Rept.}\ }\textbf {\bibinfo
  {volume} {68}},\ \bibinfo {pages} {189} (\bibinfo {year} {1981})}\BibitemShut
  {NoStop}%
\bibitem [{\citenamefont {Cremmer}\ \emph {et~al.}(1978)\citenamefont
  {Cremmer}, \citenamefont {Julia},\ and\ \citenamefont {Scherk}}]{Cre78}%
  \BibitemOpen
  \bibfield  {author} {\bibinfo {author} {\bibfnamefont {E.}~\bibnamefont
  {Cremmer}}, \bibinfo {author} {\bibfnamefont {B.}~\bibnamefont {Julia}}, \
  and\ \bibinfo {author} {\bibfnamefont {J.}~\bibnamefont {Scherk}},\ }\href
  {\doibase 10.1016/0370-2693(78)90894-8} {\bibfield  {journal} {\bibinfo
  {journal} {Phys. Lett.}\ }\textbf {\bibinfo {volume} {B76}},\ \bibinfo
  {pages} {409} (\bibinfo {year} {1978})}\BibitemShut {NoStop}%
\bibitem [{\citenamefont {Iorio}\ and\ \citenamefont
  {Lambiase}(2012)}]{Iorio11}%
  \BibitemOpen
  \bibfield  {author} {\bibinfo {author} {\bibfnamefont {A.}~\bibnamefont
  {Iorio}}\ and\ \bibinfo {author} {\bibfnamefont {G.}~\bibnamefont
  {Lambiase}},\ }\href {\doibase 10.1016/j.physletb.2012.08.023} {\bibfield
  {journal} {\bibinfo  {journal} {Phys. Lett.}\ }\textbf {\bibinfo {volume}
  {B716}},\ \bibinfo {pages} {334} (\bibinfo {year} {2012})},\ \Eprint
  {http://arxiv.org/abs/1108.2340} {arXiv:1108.2340 [cond-mat.mtrl-sci]}
  \BibitemShut {NoStop}%
\bibitem [{\citenamefont {MacDowell}\ and\ \citenamefont
  {Mansouri}(1977)}]{MacDowell:1977jt}%
  \BibitemOpen
  \bibfield  {author} {\bibinfo {author} {\bibfnamefont {S.}~\bibnamefont
  {MacDowell}}\ and\ \bibinfo {author} {\bibfnamefont {F.}~\bibnamefont
  {Mansouri}},\ }\href {\doibase 10.1103/PhysRevLett.38.1376,
  10.1103/PhysRevLett.38.739} {\bibfield  {journal} {\bibinfo  {journal}
  {Phys.Rev.Lett.}\ }\textbf {\bibinfo {volume} {38}},\ \bibinfo {pages} {739}
  (\bibinfo {year} {1977})}\BibitemShut {NoStop}%
\bibitem [{\citenamefont {Stelle}\ and\ \citenamefont
  {West}(1980)}]{Stelle:1979aj}%
  \BibitemOpen
  \bibfield  {author} {\bibinfo {author} {\bibfnamefont {K.}~\bibnamefont
  {Stelle}}\ and\ \bibinfo {author} {\bibfnamefont {P.~C.}\ \bibnamefont
  {West}},\ }\href {\doibase 10.1103/PhysRevD.21.1466} {\bibfield  {journal}
  {\bibinfo  {journal} {Phys.Rev.}\ }\textbf {\bibinfo {volume} {D21}},\
  \bibinfo {pages} {1466} (\bibinfo {year} {1980})}\BibitemShut {NoStop}%
\bibitem [{\citenamefont {McCarthy}\ and\ \citenamefont
  {Pagels}(1986)}]{McCarthy:1985nt}%
  \BibitemOpen
  \bibfield  {author} {\bibinfo {author} {\bibfnamefont {J.~G.}\ \bibnamefont
  {McCarthy}}\ and\ \bibinfo {author} {\bibfnamefont {H.~R.}\ \bibnamefont
  {Pagels}},\ }\href {\doibase 10.1016/0550-3213(86)90192-6} {\bibfield
  {journal} {\bibinfo  {journal} {Nucl.Phys.}\ }\textbf {\bibinfo {volume}
  {B266}},\ \bibinfo {pages} {687} (\bibinfo {year} {1986})}\BibitemShut
  {NoStop}%
\bibitem [{\citenamefont {Wise}(2010)}]{Wise:2006sm}%
  \BibitemOpen
  \bibfield  {author} {\bibinfo {author} {\bibfnamefont {D.~K.}\ \bibnamefont
  {Wise}},\ }\href {\doibase 10.1088/0264-9381/27/15/155010} {\bibfield
  {journal} {\bibinfo  {journal} {Class.Quant.Grav.}\ }\textbf {\bibinfo
  {volume} {27}},\ \bibinfo {pages} {155010} (\bibinfo {year} {2010})},\
  \Eprint {http://arxiv.org/abs/gr-qc/0611154} {arXiv:gr-qc/0611154 [gr-qc]}
  \BibitemShut {NoStop}%
\bibitem [{\citenamefont {Huerta}\ and\ \citenamefont {Zanelli}(1993)}]{HZ93}%
  \BibitemOpen
  \bibfield  {author} {\bibinfo {author} {\bibfnamefont {L.}~\bibnamefont
  {Huerta}}\ and\ \bibinfo {author} {\bibfnamefont {J.}~\bibnamefont
  {Zanelli}},\ }\href {\doibase 10.1103/PhysRevLett.71.3622} {\bibfield
  {journal} {\bibinfo  {journal} {Phys. Rev. Lett.}\ }\textbf {\bibinfo
  {volume} {71}},\ \bibinfo {pages} {3622} (\bibinfo {year}
  {1993})}\BibitemShut {NoStop}%
\bibitem [{\citenamefont {Alvarez}\ \emph
  {et~al.}(2014{\natexlab{b}})\citenamefont {Alvarez}, \citenamefont {Pais},
  \citenamefont {Rodríguez}, \citenamefont {Salgado-Rebolledo},\ and\
  \citenamefont {Zanelli}}]{Alv14a}%
  \BibitemOpen
  \bibfield  {author} {\bibinfo {author} {\bibfnamefont {P.~D.}\ \bibnamefont
  {Alvarez}}, \bibinfo {author} {\bibfnamefont {P.}~\bibnamefont {Pais}},
  \bibinfo {author} {\bibfnamefont {E.}~\bibnamefont {Rodríguez}}, \bibinfo
  {author} {\bibfnamefont {P.}~\bibnamefont {Salgado-Rebolledo}}, \ and\
  \bibinfo {author} {\bibfnamefont {J.}~\bibnamefont {Zanelli}},\ }\href
  {\doibase 10.1016/j.physletb.2014.09.032} {\bibfield  {journal} {\bibinfo
  {journal} {Phys. Lett.}\ }\textbf {\bibinfo {volume} {B738}},\ \bibinfo
  {pages} {134} (\bibinfo {year} {2014}{\natexlab{b}})},\ \Eprint
  {http://arxiv.org/abs/1405.6657} {arXiv:1405.6657 [gr-qc]} \BibitemShut
  {NoStop}%
\bibitem [{\citenamefont {Bañados}\ \emph {et~al.}(1992)\citenamefont
  {Bañados}, \citenamefont {Teitelboim},\ and\ \citenamefont
  {Zanelli}}]{Ban92}%
  \BibitemOpen
  \bibfield  {author} {\bibinfo {author} {\bibfnamefont {M.}~\bibnamefont
  {Bañados}}, \bibinfo {author} {\bibfnamefont {C.}~\bibnamefont
  {Teitelboim}}, \ and\ \bibinfo {author} {\bibfnamefont {J.}~\bibnamefont
  {Zanelli}},\ }\href {\doibase 10.1103/PhysRevLett.69.1849} {\bibfield
  {journal} {\bibinfo  {journal} {Phys. Rev. Lett.}\ }\textbf {\bibinfo
  {volume} {69}},\ \bibinfo {pages} {1849} (\bibinfo {year} {1992})},\ \Eprint
  {http://arxiv.org/abs/hep-th/9204099} {arXiv:hep-th/9204099 [hep-th]}
  \BibitemShut {NoStop}%
\bibitem [{\citenamefont {Bogomolny}(1976)}]{Bogomolny:1975de}%
  \BibitemOpen
  \bibfield  {author} {\bibinfo {author} {\bibfnamefont {E.}~\bibnamefont
  {Bogomolny}},\ }\href@noop {} {\bibfield  {journal} {\bibinfo  {journal}
  {Sov.J.Nucl.Phys.}\ }\textbf {\bibinfo {volume} {24}},\ \bibinfo {pages}
  {449} (\bibinfo {year} {1976})}\BibitemShut {NoStop}%
\bibitem [{\citenamefont {Prasad}\ and\ \citenamefont
  {Sommerfield}(1975)}]{Prasad:1975kr}%
  \BibitemOpen
  \bibfield  {author} {\bibinfo {author} {\bibfnamefont {M.}~\bibnamefont
  {Prasad}}\ and\ \bibinfo {author} {\bibfnamefont {C.~M.}\ \bibnamefont
  {Sommerfield}},\ }\href {\doibase 10.1103/PhysRevLett.35.760} {\bibfield
  {journal} {\bibinfo  {journal} {Phys.Rev.Lett.}\ }\textbf {\bibinfo {volume}
  {35}},\ \bibinfo {pages} {760} (\bibinfo {year} {1975})}\BibitemShut
  {NoStop}%
\bibitem [{\citenamefont {Wallace}(1947)}]{Wallace}%
  \BibitemOpen
  \bibfield  {author} {\bibinfo {author} {\bibfnamefont {P.~R.}\ \bibnamefont
  {Wallace}},\ }\href {\doibase 10.1103/PhysRev.71.622} {\bibfield  {journal}
  {\bibinfo  {journal} {Phys. Rev.}\ }\textbf {\bibinfo {volume} {71}},\
  \bibinfo {pages} {622} (\bibinfo {year} {1947})}\BibitemShut {NoStop}%
\bibitem [{\citenamefont {Semenoff}(1984)}]{Semenoff}%
  \BibitemOpen
  \bibfield  {author} {\bibinfo {author} {\bibfnamefont {G.~W.}\ \bibnamefont
  {Semenoff}},\ }\href {\doibase 10.1103/PhysRevLett.53.2449} {\bibfield
  {journal} {\bibinfo  {journal} {Phys. Rev. Lett.}\ }\textbf {\bibinfo
  {volume} {53}},\ \bibinfo {pages} {2449} (\bibinfo {year}
  {1984})}\BibitemShut {NoStop}%
\bibitem [{\citenamefont {Novoselov}\ \emph {et~al.}(2004)\citenamefont
  {Novoselov}, \citenamefont {Geim}, \citenamefont {Morozov}, \citenamefont
  {Jiang}, \citenamefont {Zhang}, \citenamefont {Dubonos}, \citenamefont
  {Grigorieva},\ and\ \citenamefont {Firsov}}]{Novoselov2004}%
  \BibitemOpen
  \bibfield  {author} {\bibinfo {author} {\bibfnamefont {K.~S.}\ \bibnamefont
  {Novoselov}}, \bibinfo {author} {\bibfnamefont {A.~K.}\ \bibnamefont {Geim}},
  \bibinfo {author} {\bibfnamefont {S.~V.}\ \bibnamefont {Morozov}}, \bibinfo
  {author} {\bibfnamefont {D.}~\bibnamefont {Jiang}}, \bibinfo {author}
  {\bibfnamefont {Y.}~\bibnamefont {Zhang}}, \bibinfo {author} {\bibfnamefont
  {S.~V.}\ \bibnamefont {Dubonos}}, \bibinfo {author} {\bibfnamefont {I.~V.}\
  \bibnamefont {Grigorieva}}, \ and\ \bibinfo {author} {\bibfnamefont {A.~A.}\
  \bibnamefont {Firsov}},\ }\href {\doibase 10.1126/science.1102896} {\bibfield
   {journal} {\bibinfo  {journal} {Science}\ }\textbf {\bibinfo {volume}
  {306}},\ \bibinfo {pages} {666} (\bibinfo {year} {2004})},\ \Eprint
  {http://arxiv.org/abs/http://www.sciencemag.org/cgi/reprint/306/5696/666.pdf}
  {http://www.sciencemag.org/cgi/reprint/306/5696/666.pdf} \BibitemShut
  {NoStop}%
\bibitem [{\citenamefont {Krasnov}(2011)}]{Krasnov:2011pp}%
  \BibitemOpen
  \bibfield  {author} {\bibinfo {author} {\bibfnamefont {K.}~\bibnamefont
  {Krasnov}},\ }\href {\doibase 10.1103/PhysRevLett.106.251103} {\bibfield
  {journal} {\bibinfo  {journal} {Phys.Rev.Lett.}\ }\textbf {\bibinfo {volume}
  {106}},\ \bibinfo {pages} {251103} (\bibinfo {year} {2011})},\ \Eprint
  {http://arxiv.org/abs/1103.4498} {arXiv:1103.4498 [gr-qc]} \BibitemShut
  {NoStop}%
\bibitem [{\citenamefont {Torres-Gomez}\ and\ \citenamefont
  {Krasnov}(2013)}]{TorresGomez:2012sr}%
  \BibitemOpen
  \bibfield  {author} {\bibinfo {author} {\bibfnamefont {A.}~\bibnamefont
  {Torres-Gomez}}\ and\ \bibinfo {author} {\bibfnamefont {K.}~\bibnamefont
  {Krasnov}},\ }\href {\doibase 10.1142/S0217751X13501133} {\bibfield
  {journal} {\bibinfo  {journal} {Int.J.Mod.Phys.}\ }\textbf {\bibinfo {volume}
  {A28}},\ \bibinfo {pages} {1350113} (\bibinfo {year} {2013})},\ \Eprint
  {http://arxiv.org/abs/1212.3452} {arXiv:1212.3452} \BibitemShut {NoStop}%
\bibitem [{\citenamefont {Witten}(1988)}]{Witten:1988hc}%
  \BibitemOpen
  \bibfield  {author} {\bibinfo {author} {\bibfnamefont {E.}~\bibnamefont
  {Witten}},\ }\href {\doibase 10.1016/0550-3213(88)90143-5} {\bibfield
  {journal} {\bibinfo  {journal} {Nucl. Phys.}\ }\textbf {\bibinfo {volume}
  {B311}},\ \bibinfo {pages} {46} (\bibinfo {year} {1988})}\BibitemShut
  {NoStop}%
\bibitem [{\citenamefont {Weyl}(1950)}]{Weyl50}%
  \BibitemOpen
  \bibfield  {author} {\bibinfo {author} {\bibfnamefont {H.}~\bibnamefont
  {Weyl}},\ }\href {\doibase 10.1103/PhysRev.77.699} {\bibfield  {journal}
  {\bibinfo  {journal} {Phys. Rev.}\ }\textbf {\bibinfo {volume} {77}},\
  \bibinfo {pages} {699} (\bibinfo {year} {1950})}\BibitemShut {NoStop}%
\bibitem [{\citenamefont {Izaurieta}\ \emph {et~al.}(2005)\citenamefont
  {Izaurieta}, \citenamefont {Rodríguez},\ and\ \citenamefont
  {Salgado}}]{Iza05}%
  \BibitemOpen
  \bibfield  {author} {\bibinfo {author} {\bibfnamefont {F.}~\bibnamefont
  {Izaurieta}}, \bibinfo {author} {\bibfnamefont {E.}~\bibnamefont
  {Rodríguez}}, \ and\ \bibinfo {author} {\bibfnamefont {P.}~\bibnamefont
  {Salgado}},\ }\href@noop {} {\  (\bibinfo {year} {2005})},\ \Eprint
  {http://arxiv.org/abs/hep-th/0512014} {arXiv:hep-th/0512014} \BibitemShut
  {NoStop}%
\bibitem [{\citenamefont {Izaurieta}\ \emph {et~al.}(2007)\citenamefont
  {Izaurieta}, \citenamefont {Rodríguez},\ and\ \citenamefont
  {Salgado}}]{Iza06a}%
  \BibitemOpen
  \bibfield  {author} {\bibinfo {author} {\bibfnamefont {F.}~\bibnamefont
  {Izaurieta}}, \bibinfo {author} {\bibfnamefont {E.}~\bibnamefont
  {Rodríguez}}, \ and\ \bibinfo {author} {\bibfnamefont {P.}~\bibnamefont
  {Salgado}},\ }\href {\doibase 10.1007/s11005-007-0148-0} {\bibfield
  {journal} {\bibinfo  {journal} {Lett. Math. Phys.}\ }\textbf {\bibinfo
  {volume} {80}},\ \bibinfo {pages} {127} (\bibinfo {year} {2007})},\ \Eprint
  {http://arxiv.org/abs/hep-th/0603061} {arXiv:hep-th/0603061} \BibitemShut
  {NoStop}%
\bibitem [{\citenamefont {Mora}\ \emph {et~al.}(2006)\citenamefont {Mora},
  \citenamefont {Olea}, \citenamefont {Troncoso},\ and\ \citenamefont
  {Zanelli}}]{Mora:2006ka}%
  \BibitemOpen
  \bibfield  {author} {\bibinfo {author} {\bibfnamefont {P.}~\bibnamefont
  {Mora}}, \bibinfo {author} {\bibfnamefont {R.}~\bibnamefont {Olea}}, \bibinfo
  {author} {\bibfnamefont {R.}~\bibnamefont {Troncoso}}, \ and\ \bibinfo
  {author} {\bibfnamefont {J.}~\bibnamefont {Zanelli}},\ }\href {\doibase
  10.1088/1126-6708/2006/02/067} {\bibfield  {journal} {\bibinfo  {journal}
  {JHEP}\ }\textbf {\bibinfo {volume} {0602}},\ \bibinfo {pages} {067}
  (\bibinfo {year} {2006})},\ \Eprint {http://arxiv.org/abs/hep-th/0601081}
  {arXiv:hep-th/0601081 [hep-th]} \BibitemShut {NoStop}%
\bibitem [{\citenamefont {Mora}\ \emph {et~al.}(2004)\citenamefont {Mora},
  \citenamefont {Olea}, \citenamefont {Troncoso},\ and\ \citenamefont
  {Zanelli}}]{Mor04}%
  \BibitemOpen
  \bibfield  {author} {\bibinfo {author} {\bibfnamefont {P.}~\bibnamefont
  {Mora}}, \bibinfo {author} {\bibfnamefont {R.}~\bibnamefont {Olea}}, \bibinfo
  {author} {\bibfnamefont {R.}~\bibnamefont {Troncoso}}, \ and\ \bibinfo
  {author} {\bibfnamefont {J.}~\bibnamefont {Zanelli}},\ }\href {\doibase
  10.1088/1126-6708/2004/06/036} {\bibfield  {journal} {\bibinfo  {journal}
  {JHEP}\ }\textbf {\bibinfo {volume} {0406}},\ \bibinfo {pages} {036}
  (\bibinfo {year} {2004})},\ \Eprint {http://arxiv.org/abs/hep-th/0405267}
  {arXiv:hep-th/0405267 [hep-th]} \BibitemShut {NoStop}%
\bibitem [{\citenamefont {Barnich}\ and\ \citenamefont
  {Brandt}(2002)}]{Barnich:2001jy}%
  \BibitemOpen
  \bibfield  {author} {\bibinfo {author} {\bibfnamefont {G.}~\bibnamefont
  {Barnich}}\ and\ \bibinfo {author} {\bibfnamefont {F.}~\bibnamefont
  {Brandt}},\ }\href {\doibase 10.1016/S0550-3213(02)00251-1} {\bibfield
  {journal} {\bibinfo  {journal} {Nucl. Phys.}\ }\textbf {\bibinfo {volume}
  {B633}},\ \bibinfo {pages} {3} (\bibinfo {year} {2002})},\ \Eprint
  {http://arxiv.org/abs/hep-th/0111246} {arXiv:hep-th/0111246 [hep-th]}
  \BibitemShut {NoStop}%
\bibitem [{\citenamefont {Mielke}\ and\ \citenamefont
  {Baekler}(1991)}]{Mielke:1991nn}%
  \BibitemOpen
  \bibfield  {author} {\bibinfo {author} {\bibfnamefont {E.~W.}\ \bibnamefont
  {Mielke}}\ and\ \bibinfo {author} {\bibfnamefont {P.}~\bibnamefont
  {Baekler}},\ }\href {\doibase 10.1016/0375-9601(91)90715-K} {\bibfield
  {journal} {\bibinfo  {journal} {Phys.Lett.}\ }\textbf {\bibinfo {volume}
  {A156}},\ \bibinfo {pages} {399} (\bibinfo {year} {1991})}\BibitemShut
  {NoStop}%
\bibitem [{\citenamefont {Blagojevic}\ and\ \citenamefont
  {Vasilic}(2003)}]{Blagojevic:2003vn}%
  \BibitemOpen
  \bibfield  {author} {\bibinfo {author} {\bibfnamefont {M.}~\bibnamefont
  {Blagojevic}}\ and\ \bibinfo {author} {\bibfnamefont {M.}~\bibnamefont
  {Vasilic}},\ }\href {\doibase 10.1103/PhysRevD.68.104023} {\bibfield
  {journal} {\bibinfo  {journal} {Phys.Rev.}\ }\textbf {\bibinfo {volume}
  {D68}},\ \bibinfo {pages} {104023} (\bibinfo {year} {2003})},\ \Eprint
  {http://arxiv.org/abs/gr-qc/0307078} {arXiv:gr-qc/0307078 [gr-qc]}
  \BibitemShut {NoStop}%
\bibitem [{\citenamefont {Chamseddine}\ and\ \citenamefont
  {Mukhanov}(2013)}]{Chamseddine:2013hwa}%
  \BibitemOpen
  \bibfield  {author} {\bibinfo {author} {\bibfnamefont {A.~H.}\ \bibnamefont
  {Chamseddine}}\ and\ \bibinfo {author} {\bibfnamefont {V.}~\bibnamefont
  {Mukhanov}},\ }\href {\doibase 10.1007/JHEP11(2013)095} {\bibfield  {journal}
  {\bibinfo  {journal} {JHEP}\ }\textbf {\bibinfo {volume} {1311}},\ \bibinfo
  {pages} {095} (\bibinfo {year} {2013})},\ \Eprint
  {http://arxiv.org/abs/1308.3199} {arXiv:1308.3199 [hep-th]} \BibitemShut
  {NoStop}%
\bibitem [{\citenamefont {Adams}(1958)}]{Ada58}%
  \BibitemOpen
  \bibfield  {author} {\bibinfo {author} {\bibfnamefont {J.~F.}\ \bibnamefont
  {Adams}},\ }\href@noop {} {\bibfield  {journal} {\bibinfo  {journal} {Bull.
  Amer. Math. Soc.}\ }\textbf {\bibinfo {volume} {64}},\ \bibinfo {pages} {279}
  (\bibinfo {year} {1958})}\BibitemShut {NoStop}%
\bibitem [{\citenamefont {Adams}(1960)}]{Ada60}%
  \BibitemOpen
  \bibfield  {author} {\bibinfo {author} {\bibfnamefont {J.~F.}\ \bibnamefont
  {Adams}},\ }\href {http://www.jstor.org/stable/1970147} {\bibfield  {journal}
  {\bibinfo  {journal} {Annals of Mathematics}\ }\bibinfo {series} {Second
  Series},\ \textbf {\bibinfo {volume} {72}},\ \bibinfo {pages} {pp. 20}
  (\bibinfo {year} {1960})}\BibitemShut {NoStop}%
\bibitem [{\citenamefont {Bañados}\ \emph {et~al.}(1993)\citenamefont
  {Bañados}, \citenamefont {Henneaux}, \citenamefont {Teitelboim},\ and\
  \citenamefont {Zanelli}}]{Banh93b}%
  \BibitemOpen
  \bibfield  {author} {\bibinfo {author} {\bibfnamefont {M.}~\bibnamefont
  {Bañados}}, \bibinfo {author} {\bibfnamefont {M.}~\bibnamefont {Henneaux}},
  \bibinfo {author} {\bibfnamefont {C.}~\bibnamefont {Teitelboim}}, \ and\
  \bibinfo {author} {\bibfnamefont {J.}~\bibnamefont {Zanelli}},\ }\href
  {\doibase 10.1103/PhysRevD.48.1506} {\bibfield  {journal} {\bibinfo
  {journal} {Phys. Rev.}\ }\textbf {\bibinfo {volume} {D48}},\ \bibinfo {pages}
  {1506} (\bibinfo {year} {1993})},\ \Eprint
  {http://arxiv.org/abs/gr-qc/9302012} {arXiv:gr-qc/9302012 [gr-qc]}
  \BibitemShut {NoStop}%
\bibitem [{\citenamefont {Miskovic}\ and\ \citenamefont
  {Zanelli}(2009)}]{Mi09}%
  \BibitemOpen
  \bibfield  {author} {\bibinfo {author} {\bibfnamefont {O.}~\bibnamefont
  {Miskovic}}\ and\ \bibinfo {author} {\bibfnamefont {J.}~\bibnamefont
  {Zanelli}},\ }\href {\doibase 10.1103/PhysRevD.79.105011} {\bibfield
  {journal} {\bibinfo  {journal} {Phys. Rev.}\ }\textbf {\bibinfo {volume}
  {D79}},\ \bibinfo {pages} {105011} (\bibinfo {year} {2009})},\ \Eprint
  {http://arxiv.org/abs/0904.0475} {arXiv:0904.0475 [hep-th]} \BibitemShut
  {NoStop}%
\bibitem [{\citenamefont {Garcia}\ \emph {et~al.}(2003)\citenamefont {Garcia},
  \citenamefont {Hehl}, \citenamefont {Heinicke},\ and\ \citenamefont
  {Macias}}]{Garcia:2003nm}%
  \BibitemOpen
  \bibfield  {author} {\bibinfo {author} {\bibfnamefont {A.~A.}\ \bibnamefont
  {Garcia}}, \bibinfo {author} {\bibfnamefont {F.~W.}\ \bibnamefont {Hehl}},
  \bibinfo {author} {\bibfnamefont {C.}~\bibnamefont {Heinicke}}, \ and\
  \bibinfo {author} {\bibfnamefont {A.}~\bibnamefont {Macias}},\ }\href
  {\doibase 10.1103/PhysRevD.67.124016} {\bibfield  {journal} {\bibinfo
  {journal} {Phys.Rev.}\ }\textbf {\bibinfo {volume} {D67}},\ \bibinfo {pages}
  {124016} (\bibinfo {year} {2003})},\ \Eprint
  {http://arxiv.org/abs/gr-qc/0302097} {arXiv:gr-qc/0302097 [gr-qc]}
  \BibitemShut {NoStop}%
\bibitem [{\citenamefont {Mielke}\ and\ \citenamefont
  {Rincon~Maggiolo}(2003)}]{Mielke:2003xx}%
  \BibitemOpen
  \bibfield  {author} {\bibinfo {author} {\bibfnamefont {E.~W.}\ \bibnamefont
  {Mielke}}\ and\ \bibinfo {author} {\bibfnamefont {A.~A.}\ \bibnamefont
  {Rincon~Maggiolo}},\ }\href {\doibase 10.1103/PhysRevD.68.104026} {\bibfield
  {journal} {\bibinfo  {journal} {Phys.Rev.}\ }\textbf {\bibinfo {volume}
  {D68}},\ \bibinfo {pages} {104026} (\bibinfo {year} {2003})}\BibitemShut
  {NoStop}%
\bibitem [{\citenamefont {Blagojevic}\ and\ \citenamefont
  {Cvetkovic}(2006)}]{Blagojevic:2006jk}%
  \BibitemOpen
  \bibfield  {author} {\bibinfo {author} {\bibfnamefont {M.}~\bibnamefont
  {Blagojevic}}\ and\ \bibinfo {author} {\bibfnamefont {B.}~\bibnamefont
  {Cvetkovic}},\ }\href {\doibase 10.1088/0264-9381/23/14/013} {\bibfield
  {journal} {\bibinfo  {journal} {Class.Quant.Grav.}\ }\textbf {\bibinfo
  {volume} {23}},\ \bibinfo {pages} {4781} (\bibinfo {year} {2006})},\ \Eprint
  {http://arxiv.org/abs/gr-qc/0601006} {arXiv:gr-qc/0601006 [gr-qc]}
  \BibitemShut {NoStop}%
\bibitem [{\citenamefont {Ma}\ and\ \citenamefont {Zhao}(2014)}]{Ma:2013eaa}%
  \BibitemOpen
  \bibfield  {author} {\bibinfo {author} {\bibfnamefont {M.-S.}\ \bibnamefont
  {Ma}}\ and\ \bibinfo {author} {\bibfnamefont {R.}~\bibnamefont {Zhao}},\
  }\href {\doibase 10.1103/PhysRevD.89.044005} {\bibfield  {journal} {\bibinfo
  {journal} {Phys.Rev.}\ }\textbf {\bibinfo {volume} {D89}},\ \bibinfo {pages}
  {044005} (\bibinfo {year} {2014})},\ \Eprint {http://arxiv.org/abs/1310.1491}
  {arXiv:1310.1491 [gr-qc]} \BibitemShut {NoStop}%
\bibitem [{\citenamefont {Blagojevic}\ and\ \citenamefont
  {Cvetkovic}(2010)}]{Blagojevic:2010jv}%
  \BibitemOpen
  \bibfield  {author} {\bibinfo {author} {\bibfnamefont {M.}~\bibnamefont
  {Blagojevic}}\ and\ \bibinfo {author} {\bibfnamefont {B.}~\bibnamefont
  {Cvetkovic}},\ }\href {\doibase 10.1103/PhysRevD.81.124024} {\bibfield
  {journal} {\bibinfo  {journal} {Phys.Rev.}\ }\textbf {\bibinfo {volume}
  {D81}},\ \bibinfo {pages} {124024} (\bibinfo {year} {2010})},\ \Eprint
  {http://arxiv.org/abs/1003.3782} {arXiv:1003.3782 [gr-qc]} \BibitemShut
  {NoStop}%
\bibitem [{\citenamefont {Blagojević}\ \emph {et~al.}(2013)\citenamefont
  {Blagojević}, \citenamefont {Cvetković},\ and\ \citenamefont
  {Vasilić}}]{Blagojevic:2013aaa}%
  \BibitemOpen
  \bibfield  {author} {\bibinfo {author} {\bibfnamefont {M.}~\bibnamefont
  {Blagojević}}, \bibinfo {author} {\bibfnamefont {B.}~\bibnamefont
  {Cvetković}}, \ and\ \bibinfo {author} {\bibfnamefont {M.}~\bibnamefont
  {Vasilić}},\ }\href {\doibase 10.1103/PhysRevD.88.101501} {\bibfield
  {journal} {\bibinfo  {journal} {Phys.Rev.}\ }\textbf {\bibinfo {volume}
  {D88}},\ \bibinfo {pages} {101501} (\bibinfo {year} {2013})},\ \Eprint
  {http://arxiv.org/abs/1310.1412} {arXiv:1310.1412 [gr-qc]} \BibitemShut
  {NoStop}%
\bibitem [{\citenamefont {'t~Hooft}(1974)}]{'tHooft:1974qc}%
  \BibitemOpen
  \bibfield  {author} {\bibinfo {author} {\bibfnamefont {G.}~\bibnamefont
  {'t~Hooft}},\ }\href {\doibase 10.1016/0550-3213(74)90486-6} {\bibfield
  {journal} {\bibinfo  {journal} {Nucl.Phys.}\ }\textbf {\bibinfo {volume}
  {B79}},\ \bibinfo {pages} {276} (\bibinfo {year} {1974})}\BibitemShut
  {NoStop}%
\bibitem [{\citenamefont {Polyakov}(1974)}]{Polyakov:1974ek}%
  \BibitemOpen
  \bibfield  {author} {\bibinfo {author} {\bibfnamefont {A.~M.}\ \bibnamefont
  {Polyakov}},\ }\href@noop {} {\bibfield  {journal} {\bibinfo  {journal} {JETP
  Lett.}\ }\textbf {\bibinfo {volume} {20}},\ \bibinfo {pages} {194} (\bibinfo
  {year} {1974})}\BibitemShut {NoStop}%
\bibitem [{\citenamefont {Abbott}\ and\ \citenamefont
  {Deser}(1982)}]{Abbott:1982jh}%
  \BibitemOpen
  \bibfield  {author} {\bibinfo {author} {\bibfnamefont {L.}~\bibnamefont
  {Abbott}}\ and\ \bibinfo {author} {\bibfnamefont {S.}~\bibnamefont {Deser}},\
  }\href {\doibase 10.1016/0370-2693(82)90338-0} {\bibfield  {journal}
  {\bibinfo  {journal} {Phys.Lett.}\ }\textbf {\bibinfo {volume} {B116}},\
  \bibinfo {pages} {259} (\bibinfo {year} {1982})}\BibitemShut {NoStop}%
\bibitem [{\citenamefont {Witten}(1981)}]{Witten:1981me}%
  \BibitemOpen
  \bibfield  {author} {\bibinfo {author} {\bibfnamefont {E.}~\bibnamefont
  {Witten}},\ }\href {\doibase 10.1016/0550-3213(81)90021-3} {\bibfield
  {journal} {\bibinfo  {journal} {Nucl.Phys.}\ }\textbf {\bibinfo {volume}
  {B186}},\ \bibinfo {pages} {412} (\bibinfo {year} {1981})}\BibitemShut
  {NoStop}%
\end{thebibliography}%
\bibliographystyle{apsrev4-1}

\end{document}